\documentclass[twocolumn]{aastex63}
\usepackage{nicefrac}
\usepackage{multirow}

\def\figref#1{Figure~\ref{#1}} 
\def\tabref#1{Table~\ref{#1}} 
\def\eqref#1{Equation~\ref{#1}} 
\def\secref#1{Section~\ref{#1}} 

\def\check#1{\textcolor{red}{#1???}} %

\def\rxj{RX J1713.7$-$3946}

\def\hessj1731{HESS J1731$-$347}

\def\bat{{\it Swift}-BAT}
\def\swift{{\it Swift}}
\def\integral{{\it INTEGRAL}}
\def\ibis{{\it INTEGRAL}-IBIS}
\def\spi{{\it INTEGRAL}-SPI}

\def\comptel{COMPTEL}

\def\fermi{{\it Fermi}}
\def\lat{{\it Fermi}-LAT}

\def\flux{\hbox{${\rm erg}~{\rm cm}^{-2}~{\rm s}^{-1}$}}

\usepackage[]{acro}

\DeclareAcronym{grb}{
  short = GRB ,
  long  = gamma-ray burst,
}
 
\DeclareAcronym{agn}{
  short = AGN ,
  long  = active galactic nucleus ,
  long-plural-form = active galactic nuclei
}
\DeclareAcronym{fsrq}{
  short = FSRQ ,
  long  = flat spectrum radio quasar ,
}
  
\DeclareAcronym{lmxb}{
  short = LMXB ,
  long  = low mass X-ray binary ,
  long-plural-form = low mass X-ray binaries ,
}
\DeclareAcronym{hmxb}{
  short = HMXB ,
  long  = high mass X-ray binary ,
  long-plural-form = high mass X-ray binaries ,
}
  
\DeclareAcronym{ic}{
  short = IC ,
  long  = inverse Compton ,
}

\DeclareAcronym{nustar}{
  short = {\it NuSTAR} ,
  long  = Nuclear Spectroscopic Telescope Array ,
}

\DeclareAcronym{bi}{
  short = BI ,
  long  = backside illumination ,
}
\DeclareAcronym{fi}{
  short = FI ,
  long  = frontside illumination ,
}
\DeclareAcronym{fov}{
  short = FoV ,
  long  = field of view ,
}
  
\DeclareAcronym{sim}{
  short = SIM ,
  long  = Science Instrument Module ,
}
\DeclareAcronym{hetg}{
  short = HETG ,
  long  = High Energy Transmission Grating ,
}
\DeclareAcronym{letg}{
  short = LETG ,
  long  = Low Energy Transmission Grating ,
}
\DeclareAcronym{hrc}{
  short = HRC ,
  long  = High Resolution Camera ,
}
\DeclareAcronym{acis}{
  short = ACIS ,
  long  = Advanced CCD Imaging Spectrometer ,
}
\DeclareAcronym{hrma}{
  short = HRMA ,
  long  = High Resolution Mirror Assembly,
}

\DeclareAcronym{compton}{
  short = Compton ,
  long  = Compton Gamma Ray Observatory,
}
\DeclareAcronym{hst}{
  short = HST ,
  long  = Hubble Space Telescope ,
}
\DeclareAcronym{iact}{
  short = IACT ,
  long  = Imaging Atmospheric Cherenkov Telescope ,
}
\DeclareAcronym{wcd}{
  short = WCD ,
  long  = Water Cherenkov Detector ,
}

\DeclareAcronym{hawc}{
  short = HAWC ,
  long  = High-Altitude Water Cherenkov ,
}
\DeclareAcronym{cta}{
  short = CTA ,
  long  = Cherenkov Telescope Array ,
}

\DeclareAcronym{em}{
  short = EM ,
  long  = electromagnetic ,
}
\DeclareAcronym{ism}{
  short = ISM ,
  long  = interstellar medium ,
}
\DeclareAcronym{csm}{
  short = CSM ,
  long  = circumstellar medium ,
}
\DeclareAcronym{sne}{
  short = SNe ,
  long  = supernovae , 
}

\DeclareAcronym{iss}{
  short = ISS ,
  long  = International Space Station ,
}

\DeclareAcronym{uhecr}{
  short = UHECRs ,
  long  = ultra high energy cosmic rays , 
}
\DeclareAcronym{ta}{
  short = TA ,
  long  = Telescope Array , 
}
\DeclareAcronym{auger}{
  short = Auger ,
  long  = Pierre Auger Observatory , 
}
\DeclareAcronym{ams}{
  short = AMS ,
  long  = Alpha Magnetic Spectrometer , 
}
\DeclareAcronym{pamela}{
  short = PAMELA ,
  long  = Payload for Antimatter Matter Exploration and Light-nuclei Astrophysics , 
}
   
\DeclareAcronym{cmb}{
  short = CMB ,
  long  = Cosmic Microwave Background , 
}
\DeclareAcronym{sed}{
  short = SED ,
  long  = spectral energy distribution , 
}
\DeclareAcronym{mhd}{
  short = MHD ,
  long  = magnetohydrodynamical ,
}
\DeclareAcronym{dof}{
  short = dof ,
  long  = degree of freedom ,
}
\DeclareAcronym{cco}{
  short = CCO ,
  long  = central compact object ,
  first-style = default
}
\DeclareAcronym{lmc}{
  short = LMC ,
  long  = Large Magellanic Cloud ,
}
\DeclareAcronym{smc}{
  short = SMC ,
  long  = Small Magellanic Cloud ,
}

\DeclareAcronym{hess}{
  short = H.E.S.S. ,
  long  = High Energy Spectroscopic System ,
  first-style = default
}
\DeclareAcronym{snr}{
  short = SNR ,
  long  = supernova remnant ,
}
\DeclareAcronym{pwn}{
  short = PWN ,
  short-plural = e ,
  long  = pulsar wind nebula ,
  long-plural  = e ,
}

\DeclareAcronym{sn}{
  short = SN ,
  short-plural = e ,
  long  = supernova ,
  long-plural  = e ,
  first-style = default
}

\DeclareAcronym{nw}{
  short = NW ,
  long  = northwest ,
  first-style = default
}
\DeclareAcronym{hxc}{
  short = HXC ,
  long  = hard X-ray component ,
  first-style = default
}

\DeclareAcronym{cr}{
  short = CR ,
  long  = cosmic ray ,
}
\DeclareAcronym{psf}{
  short = PSF ,
  long  = point spread function ,
}
\DeclareAcronym{hpd}{
  short = HPD ,
  long  = half power diameter ,
}
\DeclareAcronym{fwhm}{
  short = FWHM ,
  long  = full width of half maximum ,
}

\DeclareAcronym{pic}{
  short = PIC ,
  long  = particle-in-cell ,
  tag = numerical ,
}
\DeclareAcronym{cxb}{
  short = CXB ,
  long  = Cosmic X-ray Background ,
}
\DeclareAcronym{grxe}{
  short = GRXE ,
  long  = Galactic Ridge X-ray Emission ,
}
\DeclareAcronym{pa}{
  short = PA ,
  long  = Positional Angle ,
}
\DeclareAcronym{dsa}{
  short = DSA ,
  long  = diffusive shock acceleration ,
}

\def\degr{\hbox{$^\circ$}}
\def\arcmin{\hbox{$^\prime$}}

\def\utw{\smash{\rlap{\lower5pt\hbox{$\sim$}}}}
\def\udtw{\smash{\rlap{\lower6pt\hbox{$\approx$}}}}
\def\apjl{ApJ}%
\def\apjs{ApJS}%
\def\ao{Appl.~Opt.}%
\def\aap{A\&A}%
\def\rsep{$r_{\rm sep}$}

\shorttitle{Catalog cross-match}
\shortauthors{Tsuji et al.}

\begin{document}

\title{Cross-match between the latest Swift-BAT and Fermi-LAT catalogs}

\correspondingauthor{Naomi Tsuji}
\email{naomi.tsuji@riken.jp}

\author[0000-0002-0786-7307]{Naomi Tsuji}
\affiliation{Interdisciplinary Theoretical \& Mathematical Science Program (iTHEMS), RIKEN, 2-1 Hirosawa, Wako, Saitama 351-0198, Japan}

\author{Hiroki Yoneda}
\affiliation{RIKEN, Nishina Center, 2-1 Hirosawa, Wako, Saitama 351-0198, Japan}

\author[0000-0002-7272-1136]{Yoshiyuki Inoue}
\affiliation{Department of Earth and Space Science, Graduate School of Science, Osaka University, Toyonaka, Osaka 560-0043, Japan}
\affiliation{Interdisciplinary Theoretical \& Mathematical Science Program (iTHEMS), RIKEN, 2-1 Hirosawa, Saitama 351-0198, Japan}
\affiliation{Kavli Institute for the Physics and Mathematics of the Universe (WPI), The University of Tokyo, Kashiwa 277-8583, Japan}

\author{Tsuguo Aramaki}
\affiliation{Northeastern University, 360 Huntington Ave, Boston, MA 02115, USA}

\author{Georgia Karagiorgi}
\affiliation{Columbia University, New York, NY, 10027, USA}

\author{Reshmi Mukherjee}
\affiliation{Columbia University, New York, NY, 10027, USA}

\author{Hirokazu Odaka}
\affiliation{Department of Physics, The University of Tokyo, 7-3-1 Hongo, Bunkyo, Tokyo 113-0033, Japan}
\affiliation{Kavli Institute for the Physics and Mathematics of the Universe (WPI), The University of Tokyo, Kashiwa 277-8583, Japan}




\begin{abstract}

We report the results of a cross-match study between the hard X-ray and GeV gamma-ray catalogs, by making use of the latest 105-month \bat\ and 10-yr \lat\ catalogs, respectively.
The spatial cross-matching between the two catalogs results in the matching of 132 point-like sources, including $\sim$5\% of false-match sources.
Additionally, 24 sources that have been identified as the same identifications are matched. 
Among the 75 extended sources in the \lat\ catalog, 31 sources have spatial coincidences with at least one \bat\ source inside their extent.
All the matched sources 
consist of blazars ($>60$\%), pulsars and pulsar wind nebulae ($\sim$13\%), radio galaxies ($\sim 7$\%), binaries ($\sim 5$\%), and others.
Compared to the original catalogs, the matched sources are characterized by a double-peaked photon index distribution, higher flux, and larger gamma-ray variability index.
This difference arises from the different populations of sources, particularly the large proportion of blazars (i.e., FSRQ and BL Lac).
We also report 13 cross-matched and unidentified sources.
The matched sources in this study would be promising in the intermediate energy band between the hard X-ray and GeV gamma-ray observations, that is the unexplored MeV gamma-ray domain.

\end{abstract}

\keywords{
catalogs --- 
X-rays: general ---
gamma rays: general ---
galaxies: active
}


\section{Introduction} \label{sec:intro}
The sky in the MeV gamma-ray energy range has remained unexplored for almost 30 years since the first devoted MeV detector, the Imaging Compton Telescope \comptel\ onboard the Compton Gamma-Ray Observatory (CGRO) mission \citep{Comptel} launched in April 5, 1991, was in operation.
However, there are promising discoveries to be made in this energy band \citep{Takahashi2013APh....43..142T}, which is the main motivation for sensitive and improved observations in the next decades. 
While MeV observations await the next-generation instruments, the neighboring energy bands, the hard X-ray and the GeV gamma ray, have been well studied for the last decade by, for example, \swift /Burst Alert Telescope (BAT) \citep{Barthelmy2005} and \fermi /Large Area Telescope (LAT) \citep{Fermi2009_instrument}, respectively.
These two observatories provide us with a legacy of observational data, including source catalogs, in the corresponding energy channels.
Therefore, by using the latest \bat\ and \lat\ catalogs, one can perform catalog cross-match and somewhat predict the currently unavailable information in the MeV band.

The importance of the catalog cross-match is to list promising objects in the MeV gamma-ray band.
Sources that have been detected both in the hard X-ray and GeV gamma ray would be plausible MeV gamma-ray emitting sources unless the X-ray and gamma-ray photon indices are extremely soft and hard, respectively.
This new catalog of the cross-matched sources is useful for ongoing projects for the MeV observations \citep[e.g.,][]{eASTROGAM2018,AMEGO2019,COSI_SMEX,Aramaki2020}.

The cross-match between the hard X-ray and GeV gamma-ray catalogs is also meaningful in high energy astrophysics.
Both these energy ranges point to non-thermal radiation processes, as we expect that the thermal X-ray emission does not have a substantial contribution to the hard X-ray. 
Thus, the hard X-rays originate from synchrotron radiation or \ac{ic} scattering from accelerated electrons,
while the gamma-rays are produced by a leptonic process (i.e., \ac{ic} scattering from high-energy electrons) or a hadronic process (e.g., hadronuclear interaction).
An alternative is nonthermal bremsstrahlung from accelerated particles.
If a source emits both the hard X-rays and GeV gamma rays that originate from accelerated particles (electrons or protons) via the same or different radiation mechanisms, the broadband energy spectrum gives us an important clue to understand the particle acceleration and/or the emission mechanisms.

\cite{Maselli2011} previously performed a catalog cross correlation using the 54-month \bat\ catalog (2PBC; \cite{2pbc}) and the 1-yr \lat\ catalog (1FGL; \cite{1fgl}), which had 1256 and 1451 entries, respectively.
In this paper, we revisit to the cross-matching by making use of the latest catalogs; the 105-month \bat\ catalog \citep{Oh2018} and the 10-yr \lat\ catalog (4FGL-DR2; \cite{4fgldr2}).
With the more accumulated data and better flux sensitivity, the number of sources in the latest catalogs were improved.
Both catalogs were based on the observational data of all sky surveys. 
\integral\ \citep{Winkler2003} also performed hard X-ray observations and provided us with a hard X-ray catalog \citep[see, e.g.,][]{Bird2016}. 
However, because of its non-uniform exposure toward the sky (e.g., \integral\ has deeper exposure on the Galactic plane), we complementarily use the \integral\ catalog in this study.  

In this work, we present a catalog cross-match using the latest \bat\ and \lat\ catalogs.
\secref{sec:catalog} briefly summarizes the two catalogs.
The matching method is given in \secref{sec:method}.
The results of the matched sources are presented in \secref{sec:results}.
In \secref{sec:discussion}, we compare the matched catalog with other existing catalogs in the energy bands from hard X-ray to MeV gamma ray, investigate properties of the matched sources, and discuss the unidentified sources.
The conclusions are presented in \secref{sec:conclusions}.

\section{Catalogs} \label{sec:catalog}
This work makes use of the \bat\ 105-month \citep{Oh2018} and the \lat\ fourth (Data Release-2) \citep{4fgl,4fgldr2} catalogs of hard X-ray and GeV gamma-ray sources, respectively.

\subsection{\bat\ 105-month catalog}
The Neil Gehrels Swift Observatory (\swift) started its operation after the spacecraft was launched on November 20, 2004 \citep{Gehrels2004}.
There are three scientific instruments onboard, UV/Optical Telescope (UVOT; 170--650 nm), X-ray Telescope (XRT; 0.2--10 keV), and Burst Alert Telescope (BAT; 14--195 keV). 
BAT consists of a coded-aperture mask and a large-area solid state detector (CdZnTe) array, enabling us to detect hard X-rays in the 15--150 keV energy band with a large \ac{fov} of 1.4 sr and a \ac{psf} of 17\arcmin\ \citep{Barthelmy2005}.

Although BAT is primarily designed for detecting \acp{grb}, the accumulated data allows the BAT team to perform a uniform all-sky survey and produce a hard X-ray source catalog.
The latest catalog, the \bat\ 105-month catalog \citep{Oh2018}, made use of data taken from December of 2004 to August of 2013.
Using the 105-month data, the all sky in the 14--195 keV band was uniformly covered with sensitivities of $8.40 \times 10^{-12}$~\flux\ and $7.24 \times 10^{-12}$~\flux\ for over 90\% and 50\% of the sky, respectively.
This resulted in detection of 1632 sources at $>4.8 \sigma$.
Images, 8-channel energy spectra, and month-scale light curves of the sources in the catalog are available\footnote{\url{https://swift.gsfc.nasa.gov/results/bs105mon/}}.
In the \bat\ 105-month catalog, the largest proportion is Seyfert galaxies (827 in total; including 379 Seyfert I and 448 Seyfert II), the second one is X-ray binaries (225 in total; 109 \acp{lmxb}, 108 \acp{hmxb}, and 8 others), and the third one is beamed \acp{agn} (158 in total; including \acp{fsrq} and BL Lac types (BLLs)).

\subsection{\lat\ 4FGL-DR2 catalog}

The {\it Fermi} satellite, launched on June 11, 2008, consists of two scientific instruments, Large Area Telescope (LAT) and Gamma-ray Burst Monitor (GBM).
The \lat\ is a pair-conversion gamma-ray telescope with a precision tracker and calorimeter, each consisting of a 4$\times$4 array of 16 modules, a segmented anti-coincidence detector that covers the tracker array, and a programmable trigger and data acquisition system \citep{Fermi2009_instrument}.
\lat\ enables us to perform spectroscopy in gamma-ray energies ranging from 20 MeV to more than 300 GeV with a wide \ac{fov} of 20\% of the sky.
The \ac{psf} of \lat\ is approximately 3.5\degr\ at 100 MeV and 0.1\degr\ at 10 GeV.
The other instrument, GBM, covers two thirds of the sky at a moment and detects \acp{grb} in the 8 keV--40 MeV band.

\lat\ 4th Catalog Data Release 2 (4FGL-DR2\footnote{\url{https://fermi.gsfc.nasa.gov/ssc/data/access/lat/10yr\_catalog/}}; \cite{4fgldr2}) is the latest catalog based on 10-yr observational data taken from August 4, 2008 to August 2, 2018.
The previous catalog, the 8-yr \lat\ 4th catalog (4FGL\footnote{\url{https://fermi.gsfc.nasa.gov/ssc/data/access/lat/8yr\_catalog/}}), was described in detail in \cite{4fgl}. 
These catalogs made use of the data of the all-sky survey with the flux sensitivity of $10^{-11}$--$10^{-12}$~\flux\ in the energy range of 50 MeV to 1 TeV, depending on the source location and the energy of gamma rays.
4FGL-DR2 has 5788 sources detected at $>4\sigma$, while 4FGL has 5065 sources. 
In both catalogs, 75 sources were reported to have spatial extension.
The catalogs provide us with the locations, 7-band energy spectra, and lightcurves in 2-month and 1-yr time bins\footnote{Note that 2-month lightcurves are available only in 4FGL (the 8-yr catalog).}, which are useful for cross-matching in this paper.
We mainly made use of 4FGL-DR2 for the following analyses and used 2-month lightcurves of 4FGL for reference since 4FGL-DR2 did not include 2-month lightcurves.
The three biggest source types in 4FGL-DR2 are blazars (60\%), unknown or unindentified sources (30\%), and pulsars (5\%).

Here we note that the source category defined in the \lat\ catalog has two cases, an upper case (e.g., FSRQ) and a lower case (e.g. fsrq), which respectively indicate a firm association and an association.
Throughout this paper, we also adopt the same definition for the source category of 4FGL-DR2, otherwise mentioned.




\section{cross-match -- method} \label{sec:method}

We cross-match the 1632 \bat\ sources and the 5788 \lat\ sources by a spatially matching for point-like sources (\secref{sec:method_spatial}) and extended sources (\secref{sec:method_extended}) and carry out an identification matching (\secref{sec:method_ID}).
It should be noted that we use coordinates of the detected sources, not coordinates of the associated sources, in order to calculate the angular separation between the BAT and LAT sources.

\subsection{Spatial cross-match of point sources} \label{sec:method_spatial}

\begin{figure}[ht!]
\begin{center}
\plotone{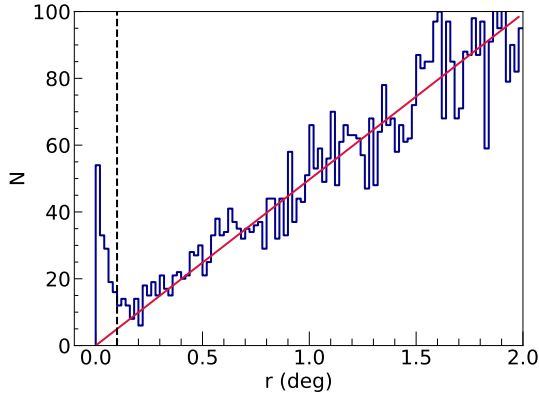}
\caption{
Distance profile in the range of $r=0$--2\degr\ with $dr=0.02$\degr. 
The red line shows the best-fit background model, with $a$ being 2500 counts~deg$^{-2}$.
The black dashed vertical line indicates the separation threshold of 0.08\degr, which suppresses the background level to 5\%.
}
\label{fig:distance_profile}
\end{center}
\end{figure}

The separation threshold for spatial cross-match (0.08\degr) was determined in the same way proposed in \cite{Itoh2020}.
First, we produced a distance profile, which is a sum of the number of the \lat\  sources located between $r$ and $r+dr$ centered at each \bat\ source as a function of the distance $r$ (\figref{fig:distance_profile}).
In \figref{fig:distance_profile}, $dr$ is set to be 0.02\degr, and the profile is generated up to $r=2.0$\degr.
The distance profile contained a spike around $r=0$\degr\ and a linear increase for $r>0.2$\degr.
The former feature indicates plausible associations, while the latter could correspond to false matches.
We thus fit the linearly increasing profile at $r>0.2$\degr\ with an empirical model of $N = a rdr$, where $a$ is a constant.
The best-fit parameter of $a$ was obtained to be 2500 counts~deg$^{-2}$.
In order to suppress the false associations (i.e., the background level) down to 5\%, we set the separation threshold, \rsep, to 0.08\degr.
Note that the background level of 10\% corresponded to \rsep\ of 0.12\degr.
We checked that the choices of $dr$ and the $r$ range for the distance profile did not have effects on determination of $a$ and \rsep.
The obtained \rsep\ (=0.08\degr) is much smaller than the \acp{psf} of the detectors and comparable with the average positional uncertainty that is 0.062\degr\ for \bat\ and 0.06\degr --0.08\degr for \lat.

Applying \rsep=0.08\degr, 132 sources were found to be cross-matched (i.e., that had counterparts within the separation).
Note that the number of the matched sources increased to 161 sources if we adjusted \rsep=0.12\degr, including possible 10\%\ false matches.
The 132 spatially matched sources are listed in \tabref{tab:point_source}, in which we show the source name, source type, position, spectral information (flux and photon index), and gamma-ray time variability index, taken from the original two catalogs.
We also show the derived separation and Flag which indicates the status of the matched source (see \secref{sec:point_source} for detail). 
The results are presented in \secref{sec:point_source}.

It should be noted that the position determination accuracy of both \bat\ and \lat\ depends on  brightness of sources.
Therefore we also carried out a spatial cross-match by setting the separation threshold to $\sigma_{\rm BAT} + \sigma_{\rm LAT}$, where $\sigma_{\rm BAT}$ and $\sigma_{\rm LAT}$ indicate the positional error of each source in the \bat\ and \lat\ catalogs, respectively.
This results in detection of 182 matched sources, which includes all the 132 spatially matched sources. 
Among the 50 sources that are missed in the spatial matching by \rsep =0.08\degr, 27 sources are matched extended sources (\secref{sec:method_extended}) or identification-matched sources (\secref{sec:method_ID}), and the remains are 7 unidentified sources and 16 false matches.

\subsection{Spatial cross-match of extended sources}  \label{sec:method_extended}
The 4FGL-DR2 catalog confirmed 75 extended sources, whose properties, including morphology, were provided in the catalog.
The source extensions range from 0.03\degr\ to 3.5\degr .
We cross-matched the two catalogs based on the assumption that the extended LAT sources had BAT sources within their extension. 
29 sources were matched with $d \leq \sigma_\gamma$, where $d$ is the angular separation between the center of the LAT source and the nearest BAT source, and $\sigma_\gamma$ is the gamma-ray spatial extent \citep[see][for details]{4fgl}.
Additional 2 sources (\ac{lmc} 30 Dor. West and HESS J1420$-$607) were matched with $d + \sigma_{\rm BAT} \leq \sigma_\gamma$, taking the positional error of the BAT source ($\sigma_{\rm BAT}$) into consideration.
In this paper, we defined these 31 sources as extended cross-matched sources.
It is notable that MSH 15$-$52 and Crab nebula (IC component), 
which were extended sources in 4FGL-DR2, were also positionally matched in \secref{sec:method_spatial}.

\subsection{Source identification cross-match} \label{sec:method_ID}
We also used an identification matching method to cross-match sources. 
When we cross-matched by the source names provided in the catalogs, 123 were matched between the \bat\ and \lat\ catalogs.
94 of these 123 sources were already included in the spatial match of point sources (\secref{sec:method_spatial} and \tabref{tab:point_source}), and another 5 sources were already presented in the spatial match of extended sources (\secref{sec:method_extended} and \tabref{tab:extended}), so we do not include them here.
The remaining 24 sources were not contained in our method of spatial cross-match.
Among the spatially unmatched and name-matched 24 sources, 10 were spatially matched if we adopted \rsep=0.12\degr\ in \secref{sec:method_spatial}. 
The remaining 14 sources may have been positionally unmatched because they had relatively large position errors because of the faint flux and had slightly larger separation than \rsep.
The separation was remarkably large for the galactic two pulsars, PSR J1420$-$6048 and PSR J1723$-$2837, and they had large position uncertainties because of their location in a complex region on the Galactic plane.

To search for associated sources in 4FGL-DR2, the 105-month \bat\ catalog was utilized as well as the many other catalogs listed in Table~6 of \cite{4fgl}.
In fact 4FGL-DR2 included 5 sources which were registered solely from the \bat\ catalog and not from the other catalogs in Table~6 in \cite{4fgl}.
They were included in our matched catalog, No. 131, 132, and 154--156 in \tabref{tab:point_source}.
The former two were spatially matched with in \rsep=0.08\degr, while the latter three were matched by the identifications.
It should be noted that the latter three sources had small association probability, $P< 0.6$ \citep[see][for details]{4fgl}, except for SWIFT J1808.5$-$3655.

\section{Results --- Cross-matched catalog}  \label{sec:results}

The catalog of the cross-matched sources between the \bat\ and \lat\ is provided here.
The spatial cross-match resulted in 132 matched sources, while the identification cross-match resulted in 24 more matched sources.
All the 156 matched point-like sources are summarized in \tabref{tab:point_source} and discussed in \secref{sec:point_source},
and the cross-matched extended sources are listed in \tabref{tab:extended} (\secref{sec:extended}).
\secref{sec:breakdown} presents the summary of source types of the matched sources.
It should be noted that Crab (No. 116 in \tabref{tab:point_source}) has three entries in 4FGL-DR2 (i.e., emission from the Crab pulsar, synchrotron emission from the Crab nebula, and inverse Compton scattering from the Crab nebula).
In this paper, we have listed only the synchrotron component that represents the three entries, because it corresponds to the hard X-ray emission seen by BAT.

\subsection{Cross-matched point sources} \label{sec:point_source}

The obtained 156 sources in \tabref{tab:point_source} were divided into five groups:
firmly matched source (with Flag being M in \tabref{tab:point_source}), 
false-matched source (F), 
source with different source categories between the two catalogs (D),  
unidentified source or unknown association (U),
and ambiguous source (A).
Brief descriptions of each group are given in the following.

\paragraph{Matched source (Flag=M)}
The matched source was defined as a source which was identified as the same source name and the same source type between the \bat\ and \lat\ catalogs.

\paragraph{False match (Flag=F)}
The false match indicates that a spatially matched source had different identifications and different source types in the two catalogs. 
Because \rsep=0.08\degr\ was determined as the level of false matching was reduced to 5\%, the 132 spatially matched sources would contain roughly 7 falsely matched sources.
Indeed, \tabref{tab:point_source} includes 8 sources where the two associated sources are not identical in the two catalogs. 
Among the 8 sources, 3 sources were pulsars in 4FGL-DR2 but different point sources in the BAT catalog (No. 108, 109, and 114 in \tabref{tab:point_source}). 
One source was classified as a \ac{pwn} in 4FGL-DR2 but a molecular cloud in the BAT catalog (No. 117), resulting from the fact that both sources are located in the radio arc near the complex galactic center.
The rest 4 false-match sources were globular clusters in 4FGL-DR2, but the corresponding BAT sources were \acp{lmxb} in the globular clusters (No. 125--127 and 129). 
These sources were likely false-matched because (1) they were confused by the emission from the Galactic plane ($|b| < 10$\degr\ for No. 108, 109, 117, 127, and 129), 
(2) they were relatively faint and had large uncertainty in position determination accuracy (No. 125 and 126), 
or (3) they had slightly smaller separation than \rsep\ (No. 114).

\paragraph{Different source type (Flag=D)}
The different-type source is identified as a source which has the same source name, but has different source types defined in the two catalogs.
\tabref{tab:point_source} includes 11 of these sources. 
7 sources were \acp{agn} with different subclasses defined in the two catalogs: they were Seyfert galaxies in the BAT catalog, but in 4FGL-DR2 they were classified as blazar candidate of uncertain type (bcu) (No. 81 in  \tabref{tab:point_source}), radio galaxies (No. 95, 96, 149 and 150), or starburst galaxies (No. 98 and 100).
They had the different subclasses because the hard X-ray and GeV gamma-ray radiation would originate from the same AGN but from the different mechanism.
We can naturally expect such associations. The X-ray emission in Seyfert galaxies originates in \ac{agn} coronae, which do not emit intense GeV gamma-ray emission due to internal $\gamma\gamma$ annihilation \citep{Inoue2019ApJ...880...40I,Inoue2020ApJ...891L..33I}. Since Seyfert galaxies also have star-formation activity, we 
see GeV emission from some of nearby Seyfert galaxies \citep[e.g.,][]{Fermi2012ApJ...755..164A}. In radio galaxies, the X-ray emission originates in the same way as in Seyfert galaxies, while AGN jet can dominate the gamma-ray emission \citep{Kataoka2011ApJ...740...29K}.
Another different-category sources were \acp{snr} in the BAT catalog but pulsars in 4FGL-DR2 (No. 106 and 107), and they were known \acp{snr} hosting pulsars \citep[e.g.,][]{SNRcat,Araya2021,Hitomi2018_G21.5}.
1RXS J122758.8-485343 (No. 110) was classified as a CV and pulsar in the \bat\ and \lat\ catalogs, respectively. 
Although the BAT catalog labeled it as a CV, it is also known as a peculiar hard X-ray source possibly associated with the \lat\ source.
\cite{Martino2013}, based on the multiwavelength observations from the radio to gamma-ray energy bands, suggested that the system would be a gamma-ray emitting \ac{lmxb}. Despite the extensive study, the nature of source No. 110 remains undetermined, and thus we labeled this source as Flag=D.
The other source, the Galactic center (No.
83), was classified as SGR A$^\star$ (source type is
Galactic Center) in the BAT catalog and Galactic Centre
(source type is bcu) in 4FGL-DR2.

\paragraph{{Unidentified association (Flag=U)}}
There were 9 sources with unknown associations, of which the source type was unclear either in the \bat\ or \lat\ catalogs (No.58, 65, 75, 130--132, and 154--156).
It should be noted that 4DFL-DR2 has two-type definitions of uncertain sources; unidentified type (i.e., sources without any firm associations) and unknown type (i.e., low Galactic-latitude sources associated solely by the Likelihood-Ratio method \citep[see][for detail]{4fgl}).
4FGL-DR2 has 1679 unidentified sources and 115 sources of unknown type.
In this paper, we merged both types and referred to them as the unidentified sources.
These sources, with their \acp{sed}, are discussed in \secref{sec:unID}.

\paragraph{Ambiguous sources (Flag=A)}
Three sources, No. 7, 13, and 76 in \tabref{tab:point_source}, were flagged as ambiguous, although their source types were AGNs in a broad meaning (i.e., Seyfert galaxy in the BAT catalog, but bll or bcu in 4FGL-DR2). 
If the associations defined in the two catalogs are correct, these 3 sources would be false-matched. However, the separation was smaller than the accuracy of position determination, and it might be better not to conclude that they were false-matched sources.
We, therefore, left them being ambiguous sources, and they need more investigations in the future to determine if they could be false matches or AGNs with different subclass.


\subsection{Cross-matched extended sources} \label{sec:extended}

All the BAT sources located inside the 31 LAT extended sources are listed in \tabref{tab:extended}, and the angular separation for each source from the LAT source is also shown.
12 LAT sources have more than one BAT source within the extent.
It should be noted that among the 31 sources, MSH 15$-$52 was also matched by the spatial matching method (\secref{sec:method_spatial}),
and \rxj, HESS J1837$-$069, and HESS J1632$-$478 were also matched by the identification-matching (\secref{sec:method_ID}). 
Since they were extended LAT sources, they are omitted in \secref{sec:point_source} and discussed in this section.

The breakdown of the 31 matched extended sources is as follows.
In the \lat\ catalog we had 2 galaxies (\ac{smc} and \ac{lmc}) and 3 unidentified subregions of \ac{lmc} (Far West, 30 Dor West, and North of \ac{lmc}).
Although they were positionally coincident with some HMXBs and a pulsar, the extended gamma rays are not associated with these point sources, thus setting them to false matches.
Additionally, the lobes in Centaurus~A detected by LAT were also matched as Centaurus~A (radio galaxy) in BAT.
10 \acp{pwn} in 4FGL-DR2 were matched with the associated pulsars in the \bat\ catalog, which are the central compact object of those \acp{pwn}.
There were 7 extended SNRs matched in our study.
Only two of them (RX J1713.7$-$3946 and RX J0852.0$-$4622) were known associations, while the other 5 included 3 false-matches (SNR G150.3$+$04.5, Monoceros, and gamma Cygni), one unknown association (Sim 147), and one ambiguous source (SNR G337.0$-$00.1 which hosted SGR 1627$-$41 (a magnetar) and IGR J16358-4726 (a pulsar) within its extent).
Cygnus X was the only one star forming region among the matched extended sources, and within the gamma-ray extent it contained Cyg X-3 (HMXB) and 2 AGNs. This, however, was falsely matched because the extended gamma-ray emission from the star forming region did not originate from those point sources. 
Among the five matched spp\footnote{`spp' is defined as a possible SNR or PWN in 4FGL-DR2.}, 3 (HESS J1632$-$478, HESS J1813$-$178, and Kes 73) were plausible associations between SNR or PWN in gamma-ray and SNR or pulsar in X-ray.
W 41, having a star SWIFT J1834.9$-$0846 measured by BAT, could be a possible false-match source. 
We left HESS J1809$-$193 as an ambiguous source because of the association with PSR J1811$-$1925, according to the spatial coincidence reported in \cite{HESS2018_HGPS}.
Furthermore, there were 3 unidentified extended \lat\ sources (FGES J1036.3$-$5833, FGES J1409.1$-$6121, and HESS J1808$-$204), which had BAT counterparts within their extended sources radii.
As mentioned above, Sim~147 that was matched with an unknown BAT source, SWIFT J053457.91$+$282837, could be also an unidentified source.
These 4 unidentified sources will be discussed in \ref{sec:discussion}.


\subsection{Summary of the matched sources}  \label{sec:breakdown}

\begin{figure*}[ht]
\begin{center}
\plotone{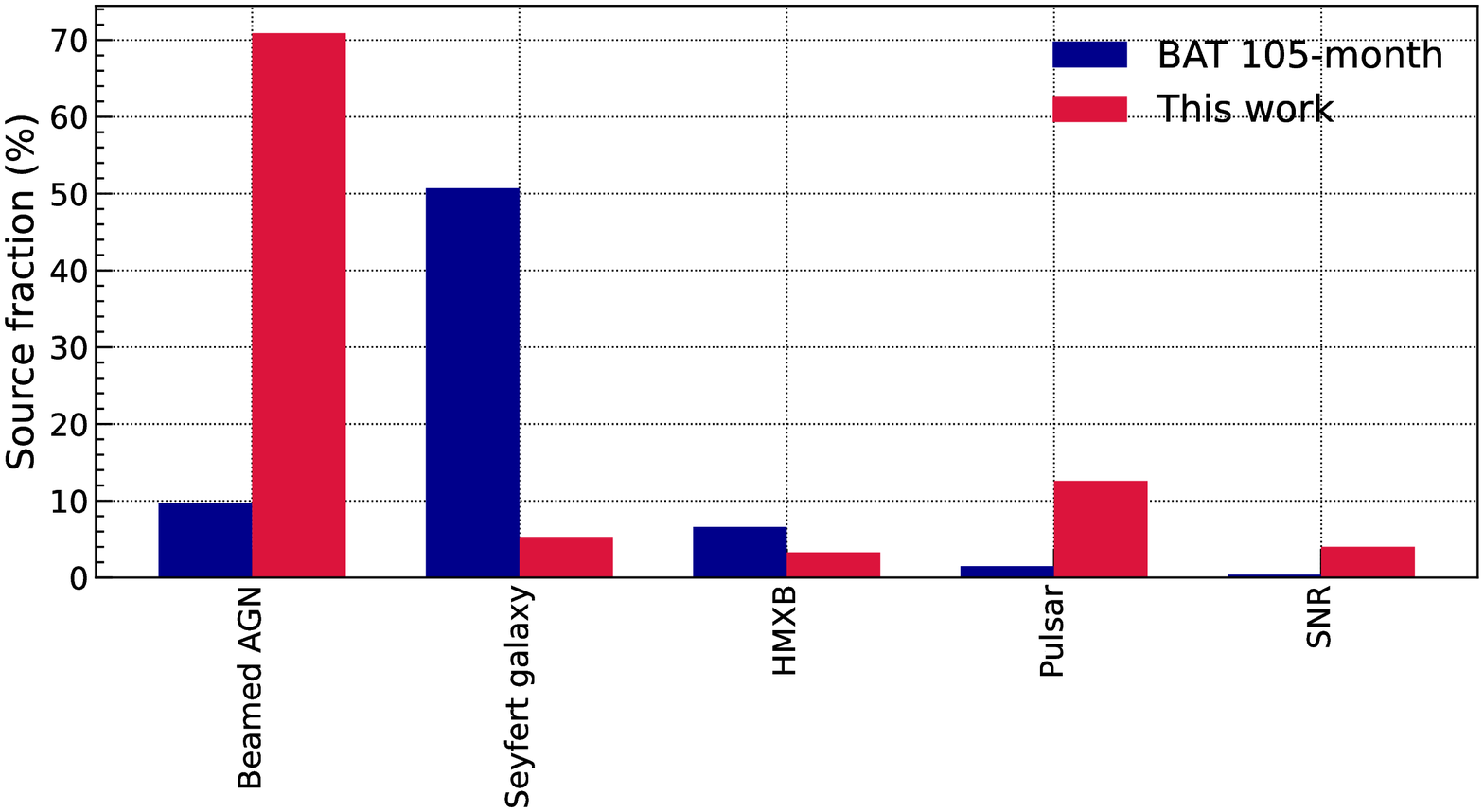}
\plotone{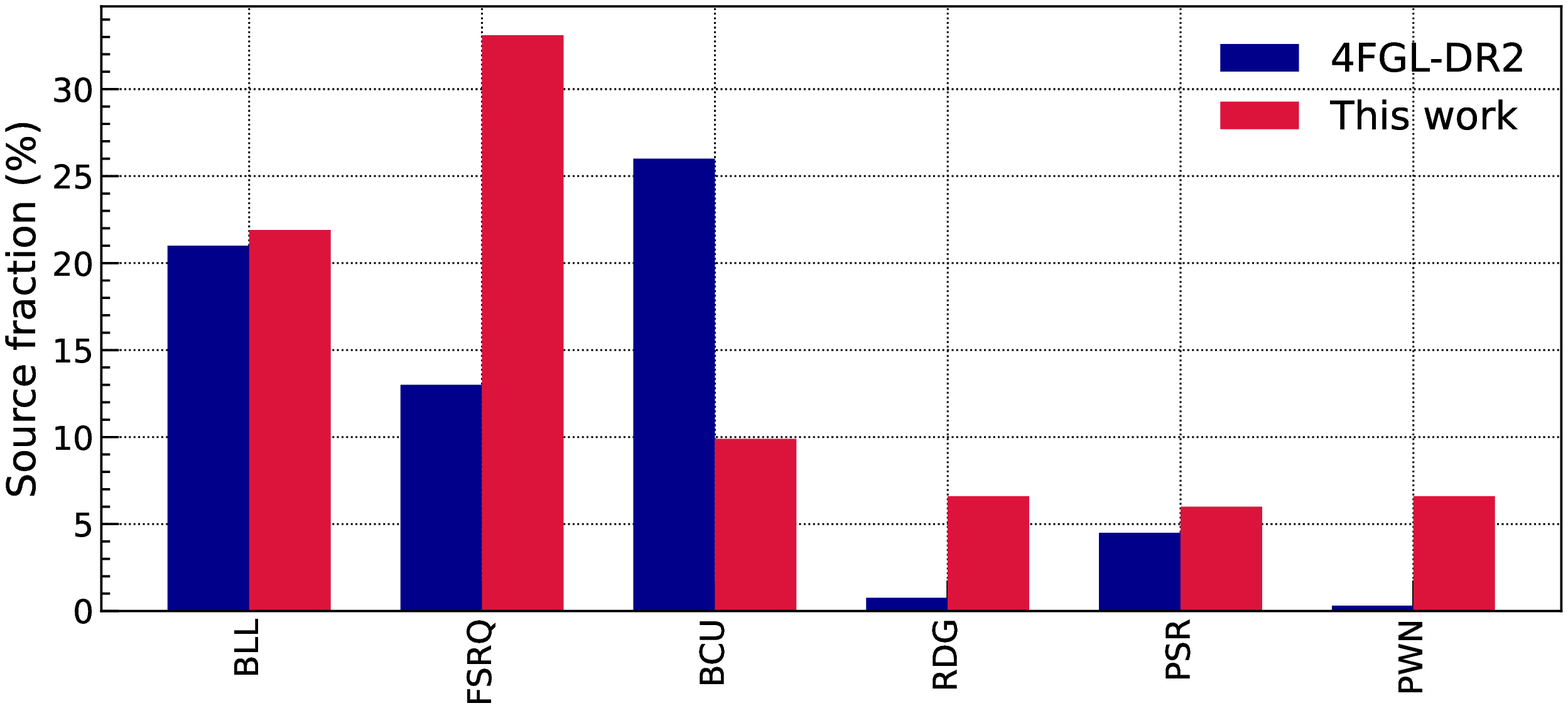}
\caption{
Top: Source type fraction of the matched catalog and the \bat\ catalog.
Bottom: Same as top for the \lat\ catalog. Note that the source category includes associations with small letters (i.e., BLL includes BLL and bll).
Only source types with the number of the matched sources of $\geq 6$ and $\geq 9$  are shown for the \bat\ and \lat\ catalogs, respectively.
}
\label{fig:summary}
\end{center}
\end{figure*}

The source type summary of the matched sources is presented in the form of the \bat\ and \lat\ definitions, respectively, in \tabref{tab:summary_bat} and \tabref{tab:summary_fermi}.
\figref{fig:summary} indicates the source type fraction of the matched sources compared to the original catalogs.
Note that only firmly matched sources (i.e., Flag is M or D in \tabref{tab:point_source} and \tabref{tab:extended}) are shown in \figref{fig:summary}.

In the \bat\ 105-month catalog, the biggest population was Seyfert galaxy, which however was not a common source category in 4FGL-DR2, resulting in a few cases of the matched Seyfert galaxies in this study. 
8 BAT Seyfert galaxies were matched, while the number reduced to 2 in the source definition of \lat.
Most of Seyfert galaxies defined in the \bat\ catalog were matched with other types of AGNs, such as bcu, radio galaxy, or starburst galaxy, as labeled as Flag=D (see \secref{sec:point_source}). 
The second largest proportion in the \bat\ catalog was X-ray binaries (HMXB, LMXB, and XRB\footnote{`XRB' in the \bat\ catalog indicates other type of X-ray binary (i.e., wind-colliding  binary system, such as Eta Carina).}).
In this work, the fraction of the matched HMXBs was roughly comparable with that of the original catalog,
although LMXBs which occupied the same fraction in the original catalog were hardly matched. 
However, the numbers of the matched HMXB and LMXB were small (i.e., five HMXBs and one LMXB), 
and thus it did not allow us further discussion about the fraction.
We note that the matched HMXBs were well known binary systems, such as LS 5039 and Cyg X-1, 
and two LMXBs classified as the unidentified sources (SAX J1808.4$-$3658 and XTE J1652$-$453) could be possible candidates of the matched sources (see \secref{sec:unID} for details).
The beamed AGNs, which were the third largest population in the original catalog, dominate in this matched catalog.
It is worth noting that the second biggest population in our catalog was pulsars, which was a minor class in the \bat\ catalog. Some of the \bat\ pulsars were matched with their nebulae in 4FGL-DR2.

In both the \lat\ and our matched catalogs, the most predominant source class was blazars. Particularly in our catalog, the fraction of BLLs was compatible with that of the original catalog, while more FSRQs were matched. This is ascribed to that FSRQs could be easily detected by \bat\ because of the typically hard spectrum in the X-ray energy range \textcolor{blue}{\citep{Toda2020ApJ...896..172T}}. The number of the matched bcu appeared small compared to the original catalog. 
In 4FGL-DR2, the number of the unidentified sources was remarkably numerous, but they were not included in our catalog. 
We found 9 cross-matched unidentified sources in total, most of which needed more investigation to confirm the association with the hard X-ray (see \secref{sec:unID} and \secref{sec:unID_extended}).
The third largest population in 4FGL-DR2 was pulsars, and we also had similar fraction of pulsars in our catalog.
It should be noted that PWNe and radio galaxies constituted a larger fraction in our catalog, while these two source categories were minor components in the original catalog.
All of the matched PWNe, however, were matched with the pulsars in the X-ray but not matched with the nebulae.

107 beamed AGNs in the \bat\ definition and 98 blazars (FSRQ, BLL, and bcu) in 4FGL-DR2 are firmly identified in our matched catalog.
These numbers were roughly consistent with that in \cite{Paliya2019}, which reported that 101 BAT blazars were gamma-ray emitting and significantly detected with \lat.
Since \cite{Paliya2019} selected the BAT blazars not based on the original definition of beamed AGN, the number of the blazars were not exactly same with our study. Indeed, 12 blazars in \cite{Paliya2019} did not appear in the our catalog.



\section{Discussion} \label{sec:discussion}

We compared our catalog to existing catalogs in the energy range from hard X-ray to sub-GeV gamma-ray, such as the \comptel\ catalog, the \integral\ catalog, the first \lat\ low energy catalog (1FLE), and the previous work by \cite{Maselli2011}, in \secref{sec:discussion_catalog_comparison}.
In \secref{sec:discussion_property}, we investigate the property of physical parameters (i.e., photon index, flux, and time variability) of our cross-matched sources.
The unidentified point-like and extended sources are discussed in \secref{sec:unID} and \secref{sec:unID_extended}, respectively.
Finally, we address the meaning of this work toward the future projects of satellites or balloon experiments in \secref{sec:future}.

\subsection{Comparison with other catalogs}    \label{sec:discussion_catalog_comparison}


\subsubsection{Comparison with \comptel\ catalog}

The \comptel\ catalog \citep{Schonfelder2000} was produced based on the first five-year data in the 0.75--30 MeV energy range.
It includes 25 steady sources, 7 line gamma-ray sources, and 31 \acp{grb}.
In this paper, we consider the 25 sources that were significantly detected at $>3\sigma$
, excluded two of them (High-velocity cloud (HVC) complexes M and A area and HVC complex C)  due to the large extent of 20--30\degr, and added 4 pulsars in Table~3 of \cite{Schonfelder2000}.
The 27 \comptel\ sources in total are shown in \tabref{tab:comptel}.

When matching with the \comptel\ catalog, the identification match was the most reasonable, and the spatial match (described in \secref{sec:method_spatial}) cannot be applicable because the coordinates of most of \comptel\ sources were taken from their counterparts.
However, the position of sources discovered by the CGRO mission (source name starting with `GRO') was determined by the \comptel\ observations. We, thus, can apply the spatial match method to these sources.

First, we conducted a name-match method to the all \comptel\ sources and searched for counterparts in the \bat\ and \lat\ catalogs.
For the identification-unmatched sources, we also picked up the nearest sources from the \bat\ and \lat\ catalogs, and then set a separation threshold of 1\degr\ for positional matching.
It should be noted that \comptel\ has the source location accuracy of $\sim$1\degr\ and the angular resolution of 3--5\degr.

The results of cross-matching are described in \tabref{tab:comptel}.
Among the 27 \comptel\ sources, 16 sources were included in our \bat\ and \lat\ cross-matching and the corresponding source No. of \tabref{tab:point_source} and \tabref{tab:extended} is given in \tabref{tab:comptel}.
The following 5 sources were matched with 4FGL-DR2 but not with the BAT catalog: PSR J0633$+$1746 (a.k.a. Geminga; No. 3 in \tabref{tab:comptel}), PSR B0656$+$14 (No.4), PSR B1055$-$52 (No. 6),  Vela/Carina (an unidentified extended emission; No. 14), and PKS 0208$-$512 (No. 22). 
The former 3 pulsars appeared faint in hard X-ray energy band.
For Nova Per 1992 (No. 12), an X-ray transient, there was no \bat\ and \lat\ counterparts.
The remaining 5 sources were ambiguous: 
GRO J2227$+$61 (No. 10),
GRO J0516$-$609 (No. 20),
GRO J1753$+$57 (No. 25),
GRO J1040$+$48 (No. 26), and 
GRO J1214$+$06 (No. 27).
Since there were no \bat\ and \lat\ counterparts within 1\degr, the position determination accuracy of \comptel, around GRO J1753$+$57 (No. 25) and GRO J1040$+$48 (No. 26), these two sources would be unmatched.
Indeed, \cite{Schonfelder2000} suggested that the emission from GRO J1753$+$57 could be modelled as a combination of emission from both GRO J1837$+$59 (a bright unidentified EGRET source) and the steep spectrum EGRET blazar QSO 1739$+$522.
GRO J2227$+$61 (No. 10) had SWIFT J2221.6$+$5952 and PSR J2229$+$6114 located 1.7\degr\ and 0.16\degr\ away from the \comptel\ emission.
GRO J0516$-$609 (No. 20) that was an unknown flaring source \citep{Bloemen1995} had a \lat\ source, PMN J0507$-$6104, within 1.03\degr.
GRO J1214$+$06 (No. 27) had two possible counterparts, 2MASX J12150077$+$0500512 and SDSS J12168$+$0541 located 0.495\degr\ and 0.567\degr\ away from the \comptel\ emission, respectively.





\subsubsection{Comparison with \ibis\ catalog}

The \integral\ observatory, launched on October 17 of 2002, consists of two main scientific instruments, the gamma-ray spectrometer SPI and the gamma-ray imager IBIS, and two sub instruments, the two X-ray monitors JEM-X and the optical monitoring camera OMC \citep{Winkler2003}.
The accumulated data taken by one of the main instruments, the coded mask telescope IBIS (particularly ISGRI, the low energy array on IBIS with a pixelated CdTe detector; \cite{Ubertini2003}), allows us a survey in the energy range from 15 keV to 1 MeV.
Using the 1000-orbit data taken from 2002 to 2010 ($\sim$110 Ms),  \cite{Bird2016} provided the 4th \ibis\ catalog, which contained 939 sources detected at $>4.5 \sigma$ in the 17--100 keV energy range. 
The latest IBIS catalog (version 43\footnote{\url{https://www.isdc.unige.ch/integral/science/catalogue}} released on September 13 of 2019) contains 1227 entries with `ISGRI\_FLAG' of $>1$, and it was used in the following.

First, we matched the latest IBIS catalog with the 105-month \bat\ catalog.
Using the same method as in \secref{sec:method_spatial} resulted in \rsep=0.26\degr, which is relatively large compared to the position uncertainty of BAT and IBIS. The large value of \rsep\ could be attributed to the fact that the distance profile of the BAT-IBIS catalog cross-match has characteristic features of a sharpened peak (i.e., the angular separation between each BAT source and the closest IBIS source is more concentrated to $r\sim0$\degr) and a low level of the background (linear increase), making the background ratio increase smoothly and \rsep\ larger. Indeed, the peak in the distance profile has a e-folding width of 0.024\degr\ in the BAT-IBIS catalog cross-match, while it is 0.082\degr\ in the BAT-LAT catalog cross-match (\figref{fig:distance_profile}).
With the separation threshold of \rsep=0.22\degr, roughly 700 sources were matched.
This indicates that we had about 900 sources detected with BAT but not with IBIS (i.e., the \bat\ catalog has roughly 1600 sources, of which 700 are also detected by \integral), and most of these sources were extragalactic, where the \bat\ had better sensitivity.
On the other hand, there were about 500 sources detected with IBIS but not with BAT, and they were distributed more on the Galactic plane, of which \integral\ had deeper exposure.
Therefore, the IBIS catalog can compensate for the sky region that has not been deeply covered by \bat.

We cross-matched the IBIS catalog and 4FGL-DR2 in the same way as described in \secref{sec:method}.
The spatial match with \rsep=0.06\degr\ resulted in 77 matched point-like sources, 
including 11 new sources that were not matched in the \bat\ and \lat\ catalog match.
Among the 11 sources, 4 were false matches, and 1 was unidentified (NVSS J175948$-$230944 in 4FGL-DR2 and IGR J17596$-$2315 in the IBIS catalog).
The remaining 6 sources were 3 FSRQs (PKS 1451$-$375, PKS 1730$-$13, PKS 1933$-$400), a bll (MS 1458.8$+$2249), an agn (PKS 1821$-$327), and a radio galaxy (M 87).
The identification match added two more sources (a radio galaxy (Can B) and an fsrq (PKS 1741$-$03)).
39 extended LAT sources were also matched,
however, including 27 sources overlapped with the \bat\ catalog in \tabref{tab:extended}, 2 false-matched sources, and 6 unidentified sources.
This led to 4 firmly matched extended sources: an SNR (IC 433), a PWN (HESS J1825$-$137), and 2 spp sources (Ken 73 and HESS J1632$-$478).
In summary, in addition to the matched sources between the \bat\ and \lat\ catalogs (\tabref{tab:point_source} and \tabref{tab:extended}), we found 8 point-like sources and 4 extended sources which were newly and firmly matched between the IBIS catalog and 4FGL-DR2.

Finally, we report on a comparison with \spi\ sources.
The \integral\ catalog contains 277 SPI sources in the 20 keV--8 MeV band (with `SPI\_FLAG' being 1) in the latest version.
29 SPI sources are matched with 4FGL-DR2 by adopting \rsep=0.06\degr, which is determined in the same way presented in \secref{sec:method_spatial}.
Among them, 26 are included in the BAT-LAT matching, one is a IBIS-LAT matched source, and the remaining two sources are false matches or ambiguous associations.

\subsubsection{Comparison with 1FLE}

\cite{Principe2018} provided the first \lat\ low energy catalog (1FLE).
This catalog was based on the 8.7-yr \lat\ data taken from August 4, 2008 to May 3, 2017 in the energy range of 30--100 MeV.
It should be noted that the \ac{psf} of even PSF3 events\footnote{Gamma rays in Pass 8 data are separated into 4 PSF event types, 0, 1, 2, and 3, where PSF0 has the largest PSF and PSF3 has the best.} is larger than 3\degr\ at $\leq$100 MeV, which is comparable with that of \comptel, 3--5\degr.
In the 1FLE catalog, 198 sources were detected at above 3$\sigma$. Among these 198 sources, 11 sources were not associated with the previous 4-yr \lat\ catalog (3FGL; \cite{3fgl}), 4FGL, and 4FGL-DR2. 

A spatial cross-match between the \bat\ 105-month catalog and 1FLE with \rsep =0.25\degr, which is comparable with the positional error of the 1FLE catalog, resulted in 19 matched point-like sources, of which 5 sources (AX J1639.0$-$4642, Mrk 766, Mrk 841, AX J1639.0$-$4642, and SWIFT J1521.6$+$3204) were not included in \tabref{tab:point_source}.
A cross-matching by the source names resulted in 35 sources being matched.
For the name-matched sources, the separation of the source coordinate between the \bat\ catalog and 1FLE was at most 1.3\degr, which is smaller than the \ac{psf} of 1FLE of $\geq 3$\degr.
Note that 14 sources are overlapped between the positionally matched sources and the name-matched sources, 
and thus the total number of point-like sources matched between the \bat\ catalog and 1FLE is 40.
In our cross-matched catalog (\tabref{tab:point_source}), we show these sources which have counterparts in 1FLE by labelling as `1FLE'.
Additionally, two extended sources, RX J1713.7$-$3946 and HESS J1632$-$478, have counterparts in 1FLE.
The BAT-1FLE matched
sources have photon indices $\lesssim 2$ in the energy band of \bat\ and $\gtrsim 3$ in the energy band of \lat\ except for Mrk 421 with  $\Gamma_{\rm BAT} >2$ and  $\Gamma_{\rm Fermi} >2$, NGC 1275 with $\Gamma_{\rm BAT} >2$, and RX J0115.7$+$2519 with $\Gamma_{\rm Fermi} <3$.
It should be noted that all the 1FLE sources matched here had associations with sources of 3FGL, and the unidentified 11 1FLE sources were not matched with the BAT sources.

\subsubsection{Comparison with Maselli et al. 2011}

In a previous study, \cite{Maselli2011} performed a catalog cross-match by using the 54-month \bat\ catalog (2PBC; 1256 sources; a flux sensitivity of (0.92--1.0)$\times 10^{-11}$~\flux ; \cite{2pbc}) and the 1-yr \lat\ catalog (1FGL; 1451 sources; a flux sensitivity of $10^{-11}$--$10^{-10}$~\flux ; \cite{1fgl}).
They reported 62 sources as firmly cross-matched sources which had the same identifications between the two catalogs.
Furthermore, 46 sources were positionally matched if the $Q$ parameter (defined as $ (r_{\rm BAT} + r_{\rm LAT}) / r_{\rm BL} $ where $r_{\rm BAT}$, $r_{\rm LAT}$, and $r_{\rm BL}$ are respectively the 
position uncertainty of a BAT source, 
that of a LAT source, and the higher value between $r_{\rm BAT}$ and $r_{\rm LAT}$) was set to be $<1.0$ \cite[see][for details]{Maselli2011}.
87 sources in total were matched by the aforementioned positional and identification matching, since 21 sources were overlapped in the two methods.
By decreasing the X-ray detection threshold to 3$\sigma$ from 4.8$\sigma$, 
the number of the hard X-ray emitting BAT sources in the direction of 1FGL sources increased to 104, which include all the 87 cross-correlated sources.

Among the firmly associated 62 sources in \cite{Maselli2011}, 8 were not included in our analysis (\tabref{tab:point_source}).
However, this discrepancy is attributed to the fact that these 8 sources were excluded either in the latest \bat\ or \lat\ catalogs.
The following 6 sources are included in 4FGL-DR2, but omitted in the latest BAT catalog probably due to flux time variation:
OI $+$280 in the \bat\ 54-month catalog (PKS 0748$+$126 in 1FGL),
RX J0948.8$+$0022 (CGRaBS J0948$+$0022),
RBS 1420 (1ES 1440$+$122),
Ap Lib,
PG 1553$+$113,
and PG 0727$-$11 (PKS 0727$-$11).
ESO 323$-$77 is in the BAT 105-month catalog, but not included in 4FGL-DR2 (\cite{Maselli2011} also mentioned that this source is a confused LAT source).
The remaining one source, 1RXS J033913.4$-$173553 (PKS 0336$-$177) had $Q>1$ (i.e., spatially unmatched) in \cite{Maselli2011}, and thus was not matched in our study.
In conclusion, all the firmly matched sources in \cite{Maselli2011} resulted in being matched in this paper, unless the sources were not excluded in the later \bat\ or \lat\ catalogs.
The number of the firmly matched sources roughly doubled in this study owing to the developed flux sensitivity of the observations, particularly that of \lat\, which was almost one order of magnitude better.


\subsection{Property of matched sources}  \label{sec:discussion_property}

In the following, we compare the photon index, flux, and time variability of the matched and unmatched sources in order to investigate the properties of the matched sources.


\begin{figure*}[ht!]
\begin{center}
\plottwo{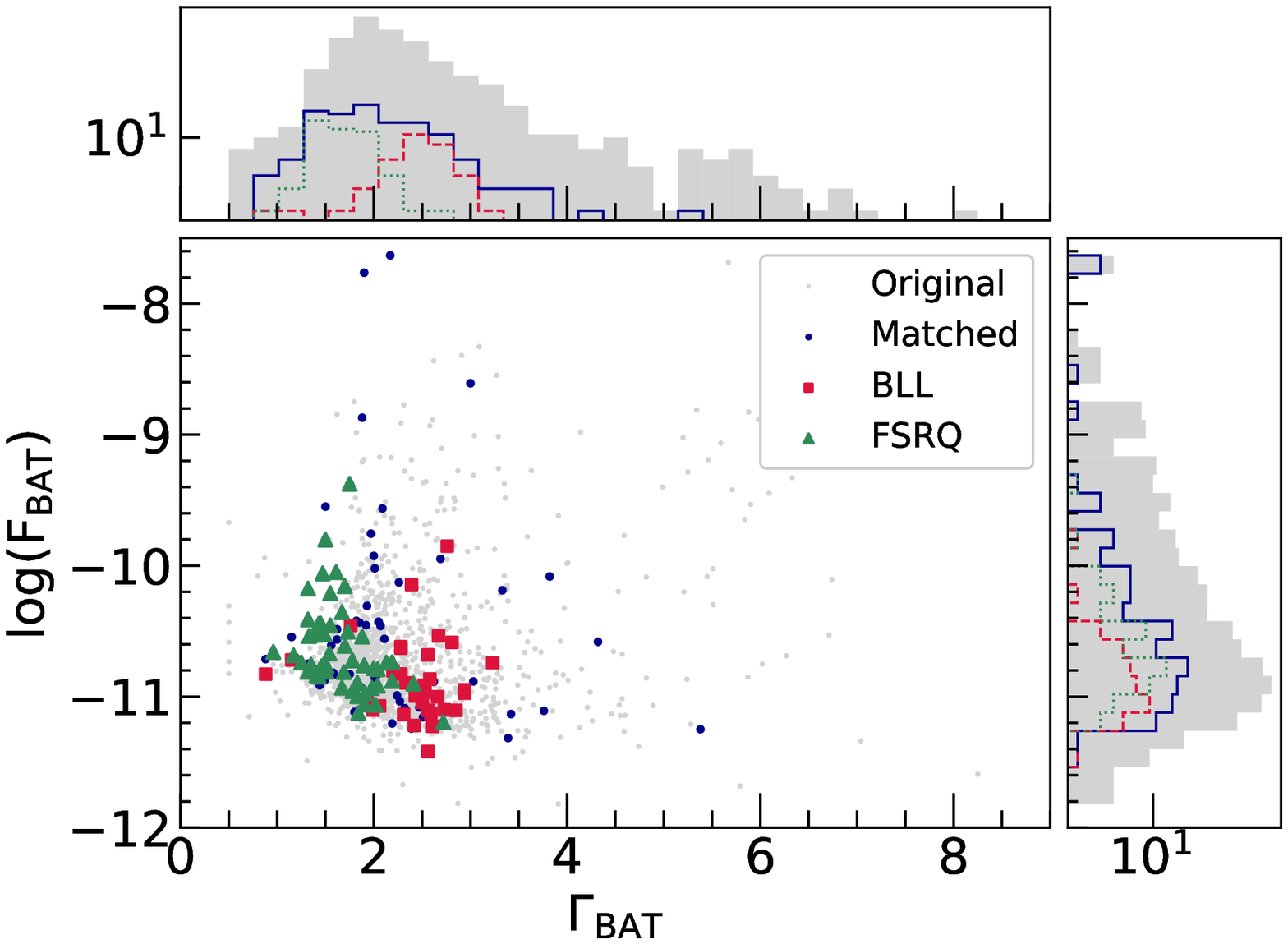}{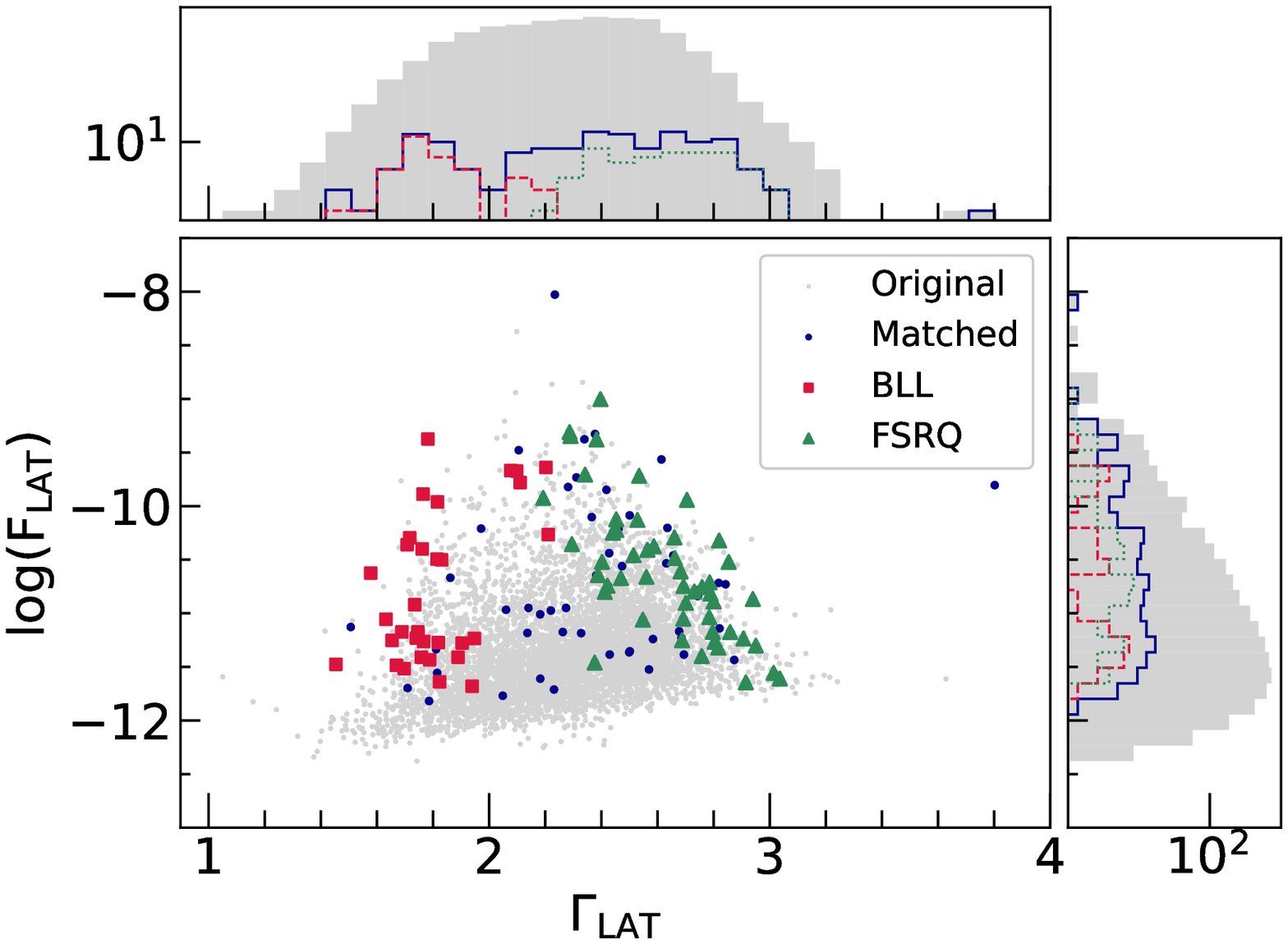}
\caption{
Correlation between $\Gamma$ and flux and their distributions of the \bat\ (left) and \lat\ (right) sources.
The grey and blue points are distributions of all the sources in the catalog and the spatially matched sources, respectively.
The distributions of the marched BLLs and FSRQs are shown in red and green, respectively.
The histograms are shown in logarithmic scale.
The figure includes the 136 sources which are firmly matched by the coordinates and the identifications, with the Flag being M or D. 
}
\label{fig:pindex_flux}
\end{center}
\end{figure*}

\figref{fig:pindex_flux} shows a correlation between a photon index ($\Gamma$) and flux and their distributions for the matched sources in this catalog and all sources in the original catalog. 
Here we used the firmly matched point-like sources (136 in total) with Flag being M or D in \tabref{tab:point_source}. 
Even when including the firmly matched extended sources, the following results did not largely change.
For the BAT sources (the left panel of \figref{fig:pindex_flux}), the distribution of $\Gamma$ for the matched sources was slightly shifted to the harder side compared to that of all sources, while the distribution of the flux was shifted to the brighter side. 
By using Kolmorogov-Smirnov (KS) statistic, we evaluated the difference of the distributions of $\Gamma$ and the flux between the matched sources and all sources in the original catalog.
The $\Gamma$ distribution showed the value of KS statistic of 0.196 and the p-value of 0.000160, which corresponded to 3.8$\sigma$, while
the flux distribution showed the value of KS statistic of 0.147 and the p-value of 0.00973, which corresponded to 2.6$\sigma$.
Hence, the distributions of $\Gamma$ and the flux had different properties at the level of $\sim 3 \sigma$.

For LAT sources (the right panel of \figref{fig:pindex_flux}), the $\Gamma$ distribution shows an apparent bimodal feature, and the distribution of the flux was clearly shifted to the brighter side, compared to all sources in the original catalog.
Similar to the aforementioned results of the BAT sources, we also found that the distributions of $\Gamma$ and the flux of the matched sources were different from those of the original catalog.
The KS statistics and the corresponding p-value were respectively 0.192 and 0.000140 ($3.8\sigma$) in the $\Gamma$ distribution, while they were respectively 0.427 and $1.49 \times 10^{-21}$ (over $5\sigma$) in the flux distribution.

The difference in the $\Gamma$ and flux properties can be explained as follows.
Among the matched point sources, the two largest populations were FSRQs (50 sources) and BLLs (33 sources).
These two classes of blazars might be part of the blazar sequence, with the synchrotron and high-energy peak at different energy bands:
in the energy range of \bat\ and \lat, FSRQs have concave-structure (i.e., hard in X-ray and soft in gamma-ray), while BLLs have convex-structure  (i.e., soft in X-ray and hard in gamma-ray).
Indeed, the double-peak feature in the $\Gamma$ distribution was ascribed to the $\Gamma$ distributions of the FSRQs and BLLs (\figref{fig:pindex_flux}).
It also should be noted that the fraction of the FSRQs in the matched sources was notably larger than that of the original catalog (\figref{fig:summary}), making the $\Gamma$ distributions modified.
The difference in the flux distributions can arise from the difference in flux sensitivity between \bat\ and \lat.
Particularly the flux distribution of the Fermi sources showed the  remarkable distinction between the matched and all sources.
The better sensitivity of \lat\ resulted in the difference in the flux distributions, recalling that the flux sensitivity of \bat\ and \lat\ are respectively $8\times 10^{-12}$~\flux\ and $\sim 1 \times 10^{-12}$~\flux. 

\begin{figure}[ht!]
\begin{center}
\plotone{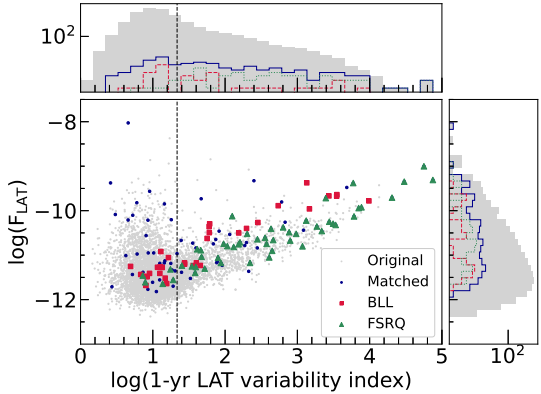}
\caption{
Same as \figref{fig:pindex_flux} for correlation between the 1-yr variability index and flux of the \lat\ sources.
The 1-yr variability index of $>$21.67, shown by the vertical dashed line, means $<$1\% chance of a steady source.
}
\label{fig:var_index}
\end{center}
\end{figure}

We also investigated property of time variation of the matched sources.
4FGL-DR2 provides us with `Variability Index', which is defined as a sum of 2$\times$Log(Likelihood) difference between flux of each time and the averaged one.
For the 10-yr lightcurve with 1-yr bin, the variability index of $>21.67$ indicates a $<1$\% chance for a steady source.
It should be noted that lightcurves with 1-yr bin and 2-month bin were available in 4FGL (the previous 8-yr \lat\ catalog), while only lightcurves with 1-yr bin were provided in 4FGL-DR2.
We made sure that the variability indices of the 1-yr and 2-month lightcurves were correlated, and the following results produced by  4FGL-DR2 were consistent with when using the corresponding variability index of the 2-month lightcurves in 4FGL.

\figref{fig:var_index} shows a correlation between the variability index and the flux and their distributions of the matched sources in our catalog and the all sources in the \lat\ catalog.
There seem to be two groups in the scatter plot in \figref{fig:var_index}: the correlated variability index and flux (i.e., the time variation can be easily detected for the brighter source) and the smaller variability index with the widely ranged flux (i.e., possible steady source).
The distribution of the variability index of the matched sources was also different from that of the original catalog, inferred from the KS statistics and p-value of 0.414 and $3.29\times 10^{-20}$ ($>5\sigma$), respectively.

Our matched sources turned out to be more variable than the sources in the original catalog.
This discrepancy arised from the fact that the matched sources mainly consisted of FSRQs and BLLs (\figref{fig:summary}), which tended to have large variability indices.
In the original catalog, 80\% of FSRQs are variable with the index of $>21.67$, and 43\% of BLLs are so. 
The difference in the distribution of the variability index could also be attributed to the fact that the brighter sources, correlated to the larger variability index, were more matched in this study.

\begin{figure*}[ht!]
\begin{center}
 \plottwo{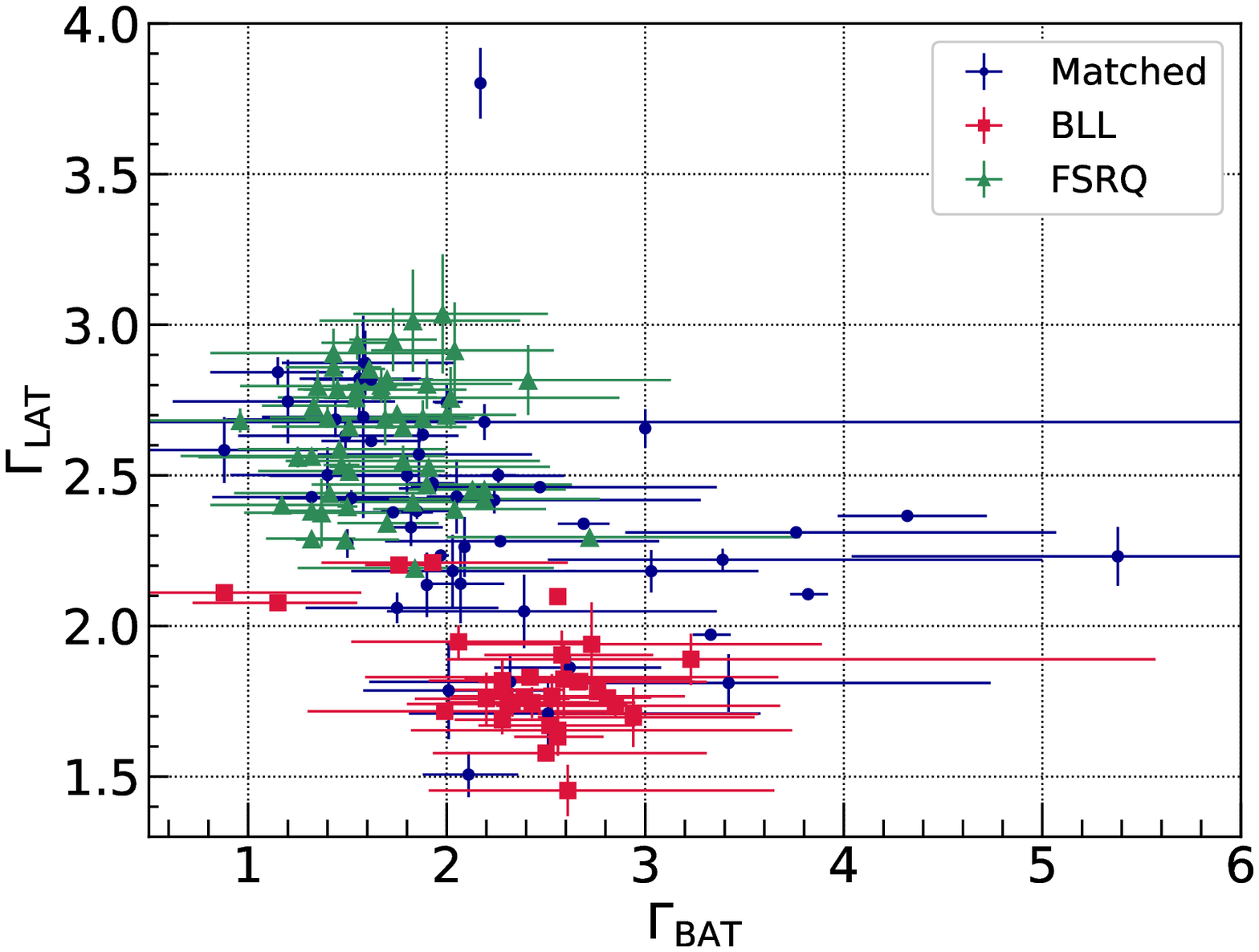}{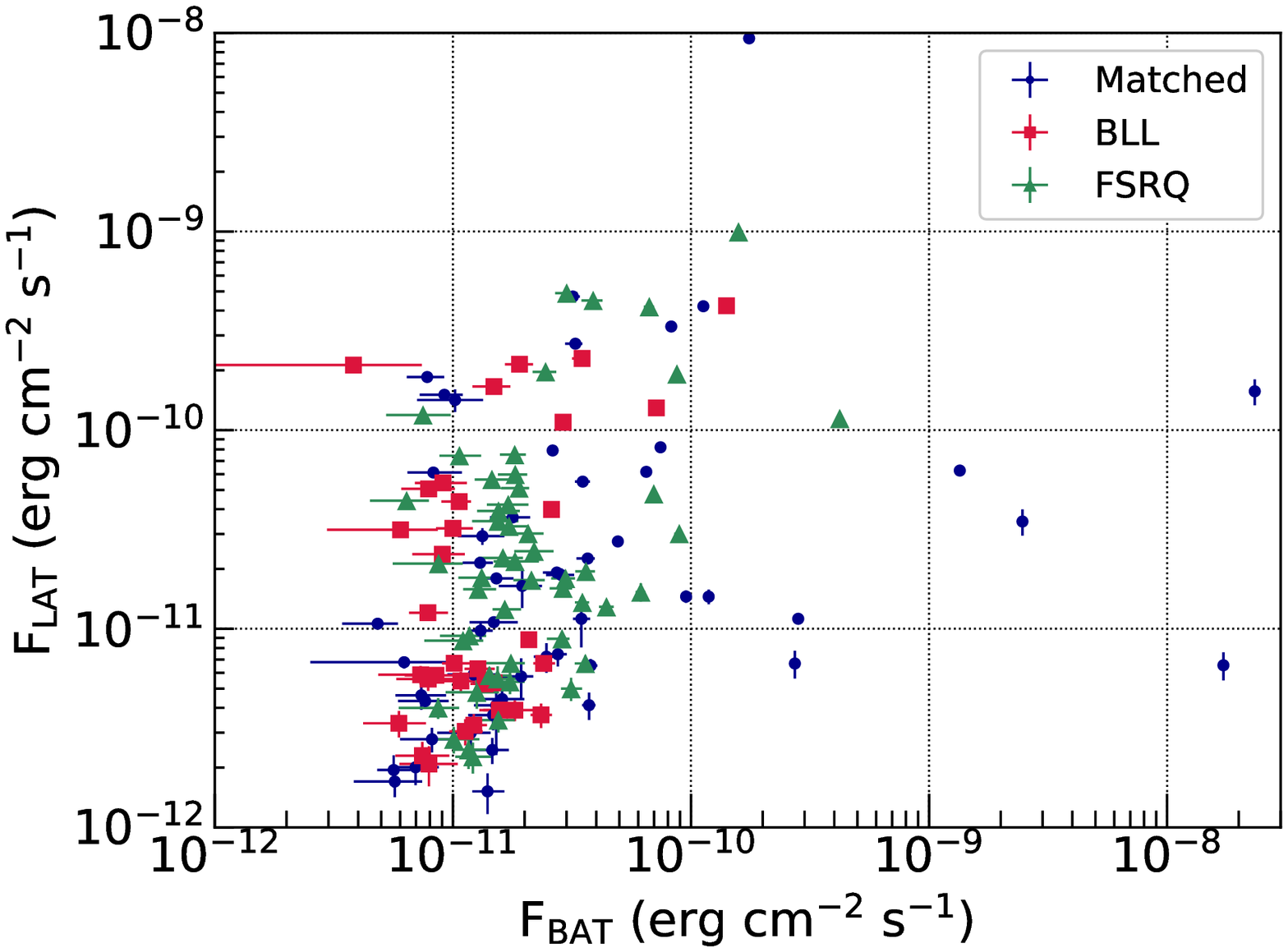}
\caption{
Scatter plots of the $\Gamma$ (left) and flux (right) of the firmly matched \bat\ and \lat\ sources.
The red and green respectively show those of BLLs and FSRQs, and the blue indicates those of the rest sources.
}
\label{fig:scatter_ff}
\end{center}
\end{figure*}

We present the correlation of the photon indices between the firmly matched \bat\ and \lat\ sources in \figref{fig:scatter_ff}.
As seen in \figref{fig:pindex_flux}, the $\Gamma_{\rm BAT}$-$\Gamma_{\rm LAT}$ diagram also confirmed two distinct populations, BLLs and FSRQs.
The right panel of \figref{fig:scatter_ff} shows the correlation of the flux of the firmly matched \bat\ and \lat\ sources.
In the hard X-ray band, the flux of the matched BLLs tends to be smaller than that of the matched FSRQs.

To summarize, \figref{fig:pindex_flux} and \figref{fig:var_index} suggest that our matched sources can be characterized by the double peak in the $\Gamma$ distribution, the higher flux, and the larger variability index, compared to the all sources in the original catalogs.
This difference would be reflected by the features of the two main populations, FSRQs and BLLs. 


\subsection{Unidentified point-like sources} \label{sec:unID}

\begin{figure*}[ht]
\begin{center}
\gridline{\fig{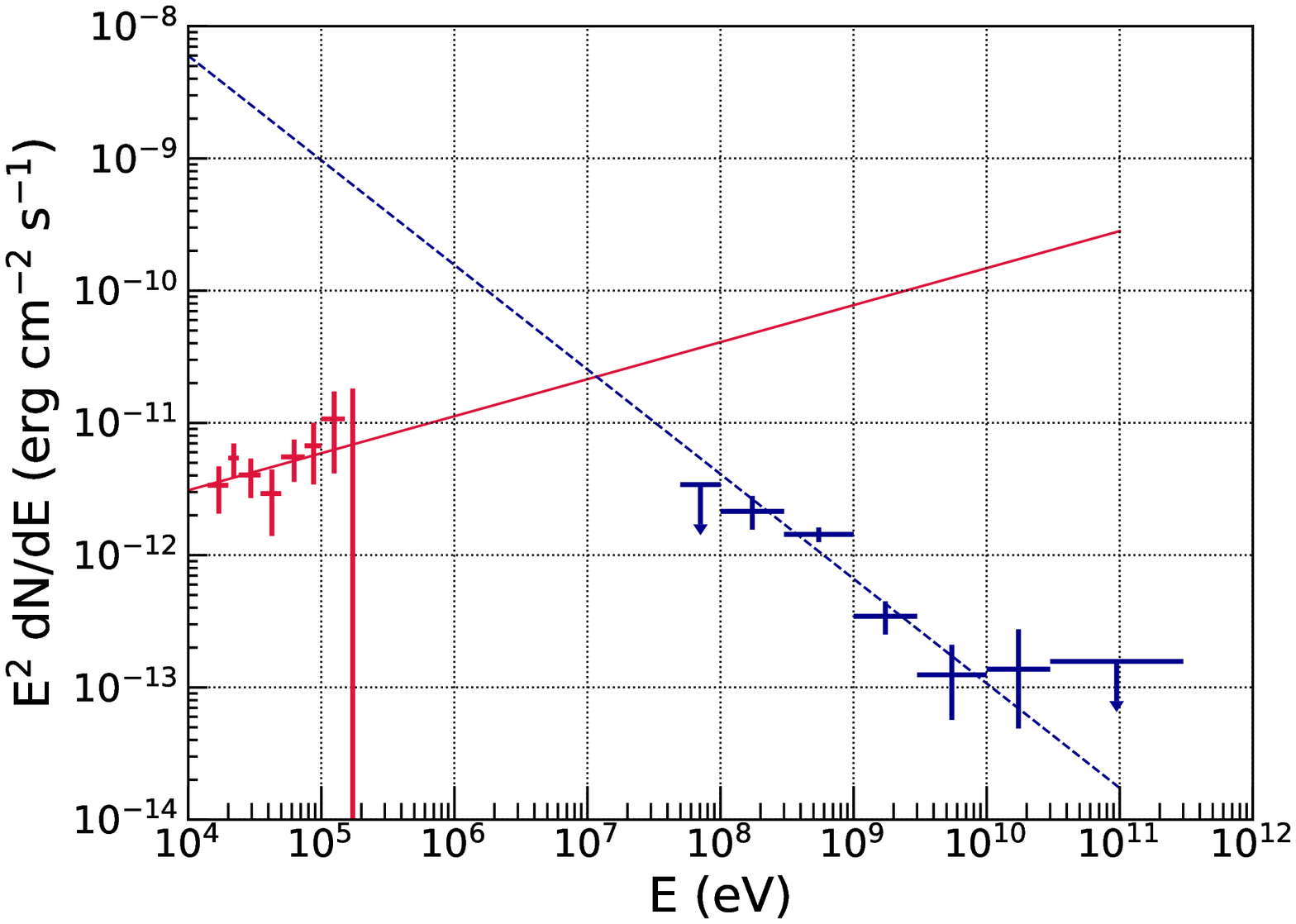}{0.32\textwidth}{SWIFT J1254.9$+$1165 (No. 58 in \tabref{tab:point_source})}
          \fig{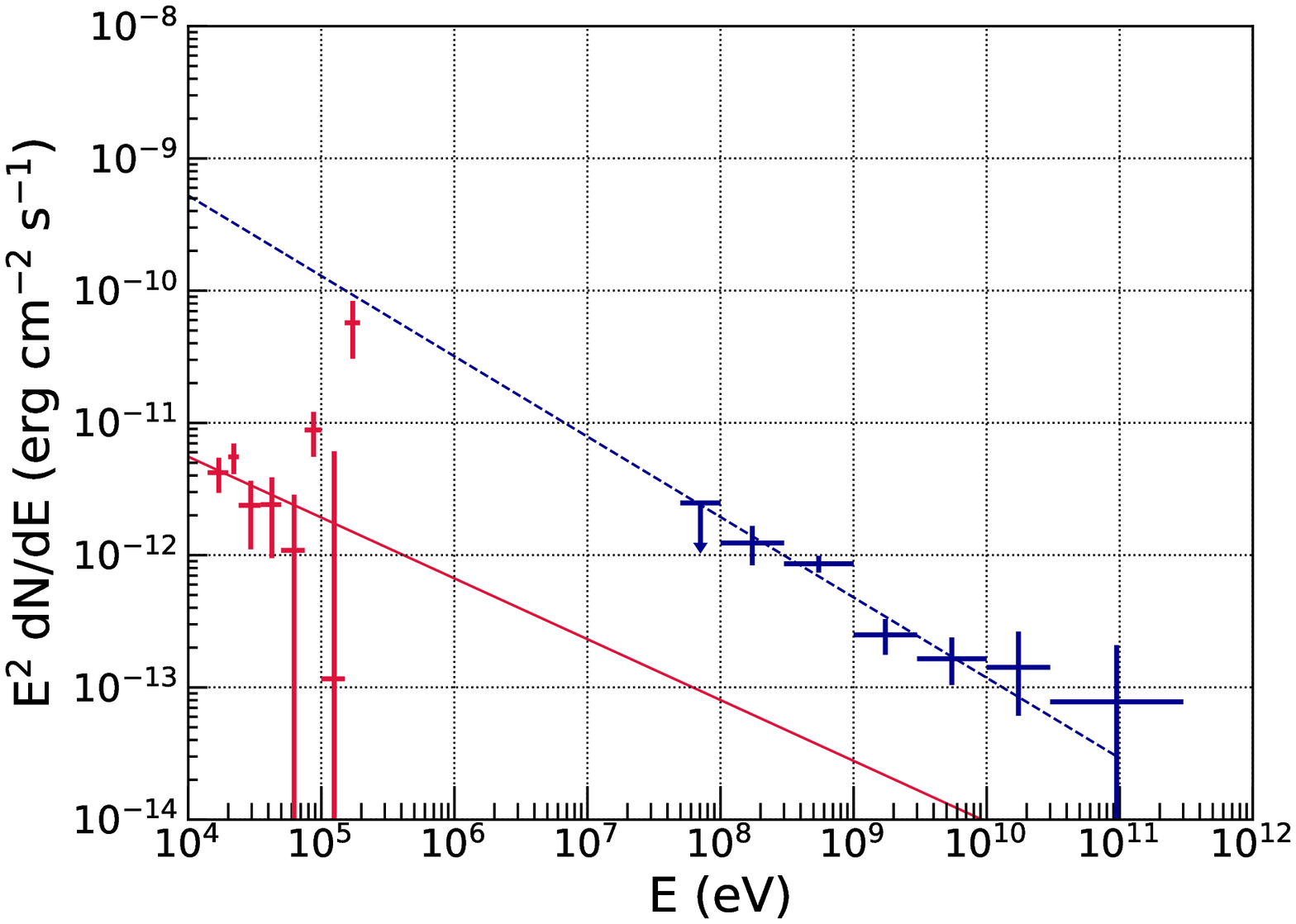}{0.32\textwidth}{SWIFT J0949.1$+$4057 (No. 65)}
          \fig{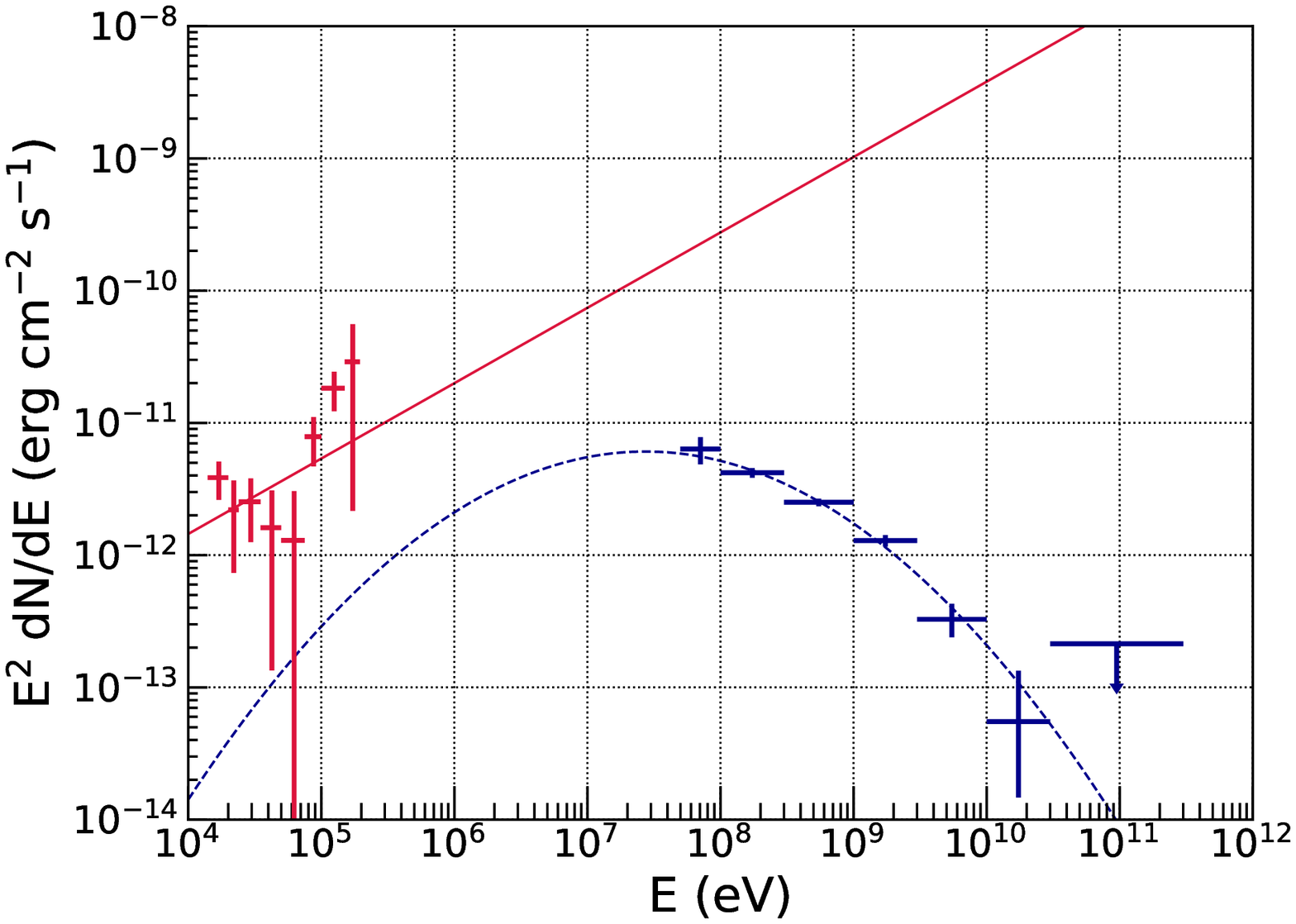}{0.32\textwidth}{PMN J0145$-$2733 (No. 75)}
          }
\gridline{\fig{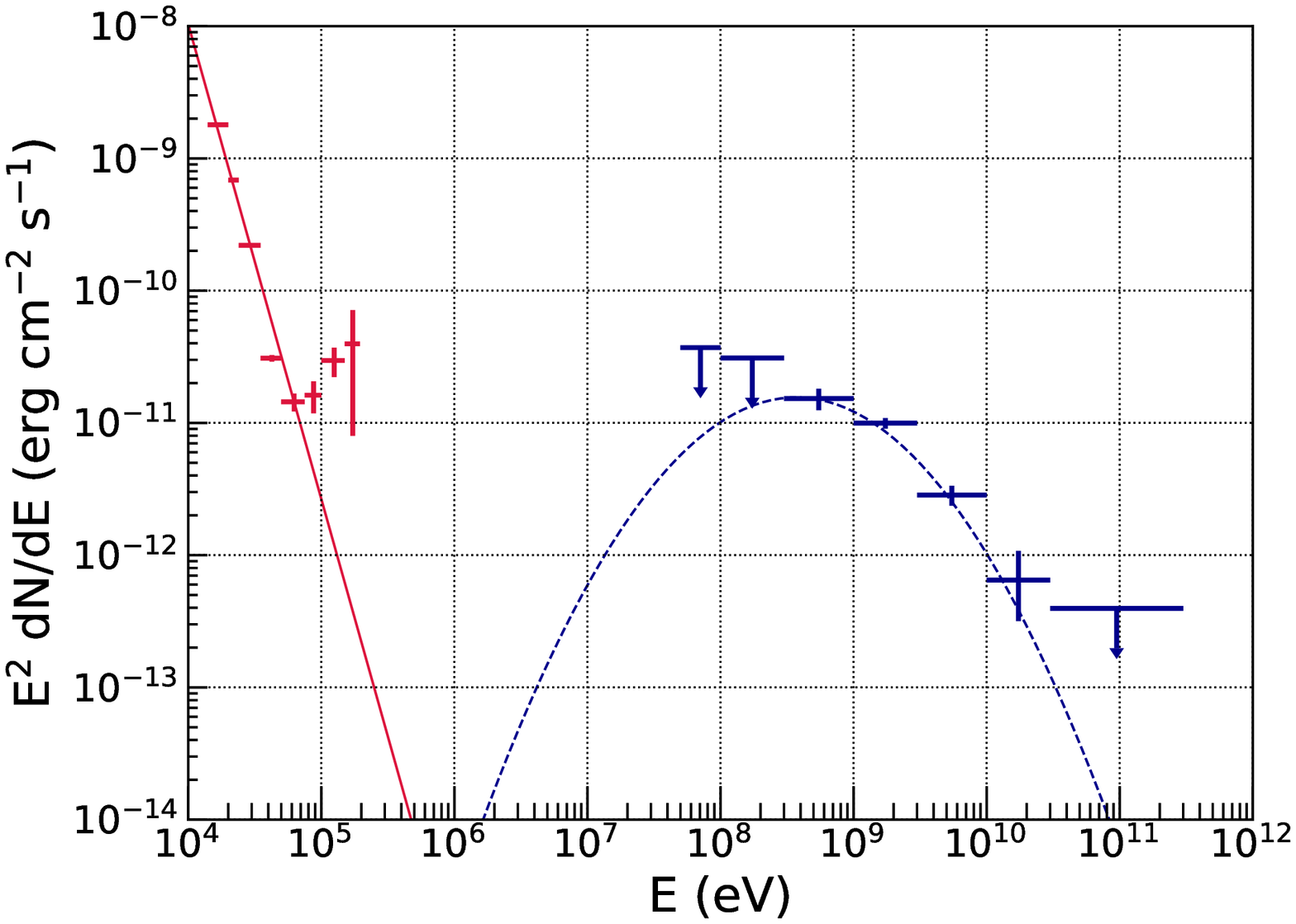}{0.32\textwidth}{GX 340$+$0 (No. 130)}
          \fig{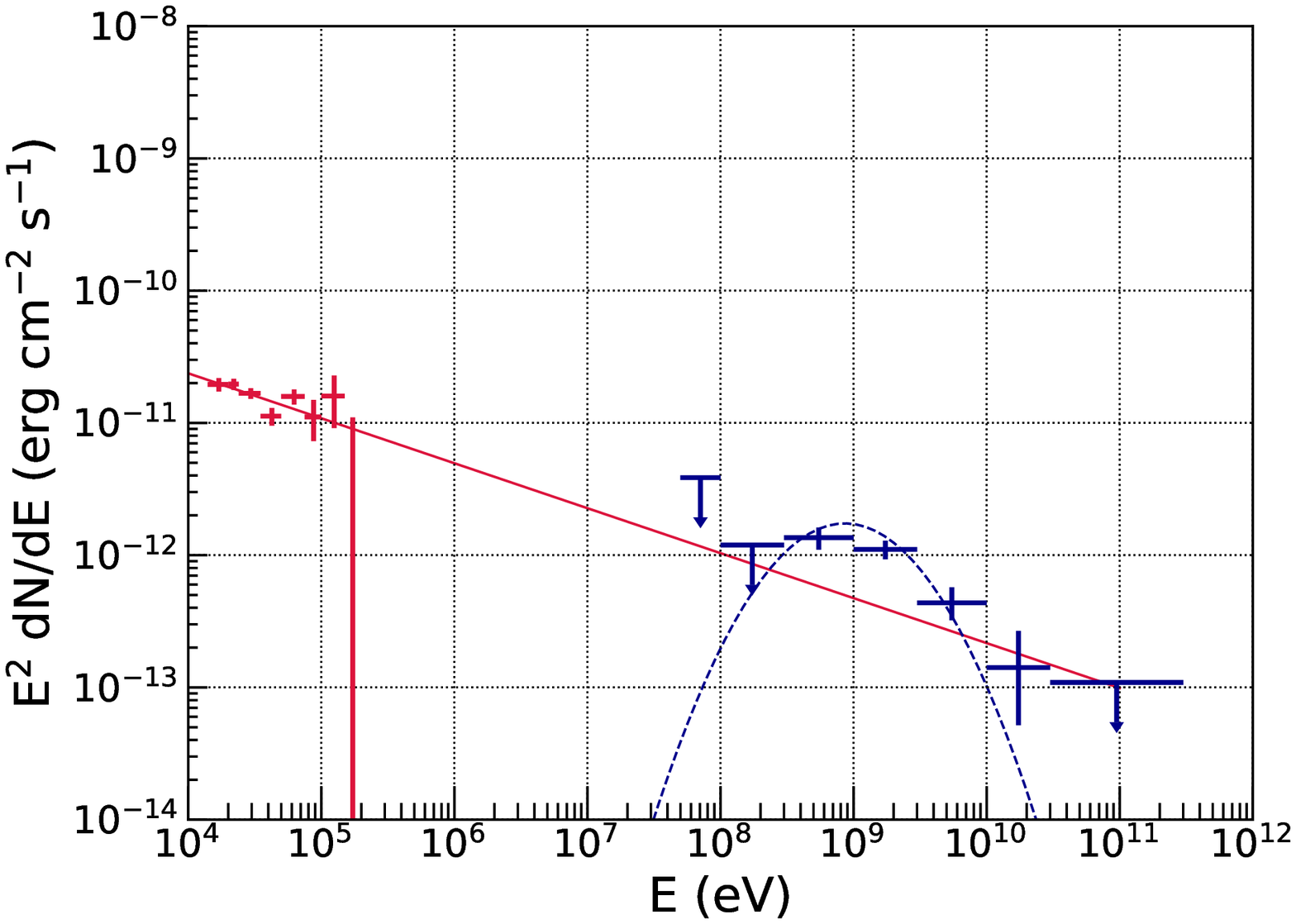}{0.32\textwidth}{SAX J1808.4$-$3658 (No. 131)}
          \fig{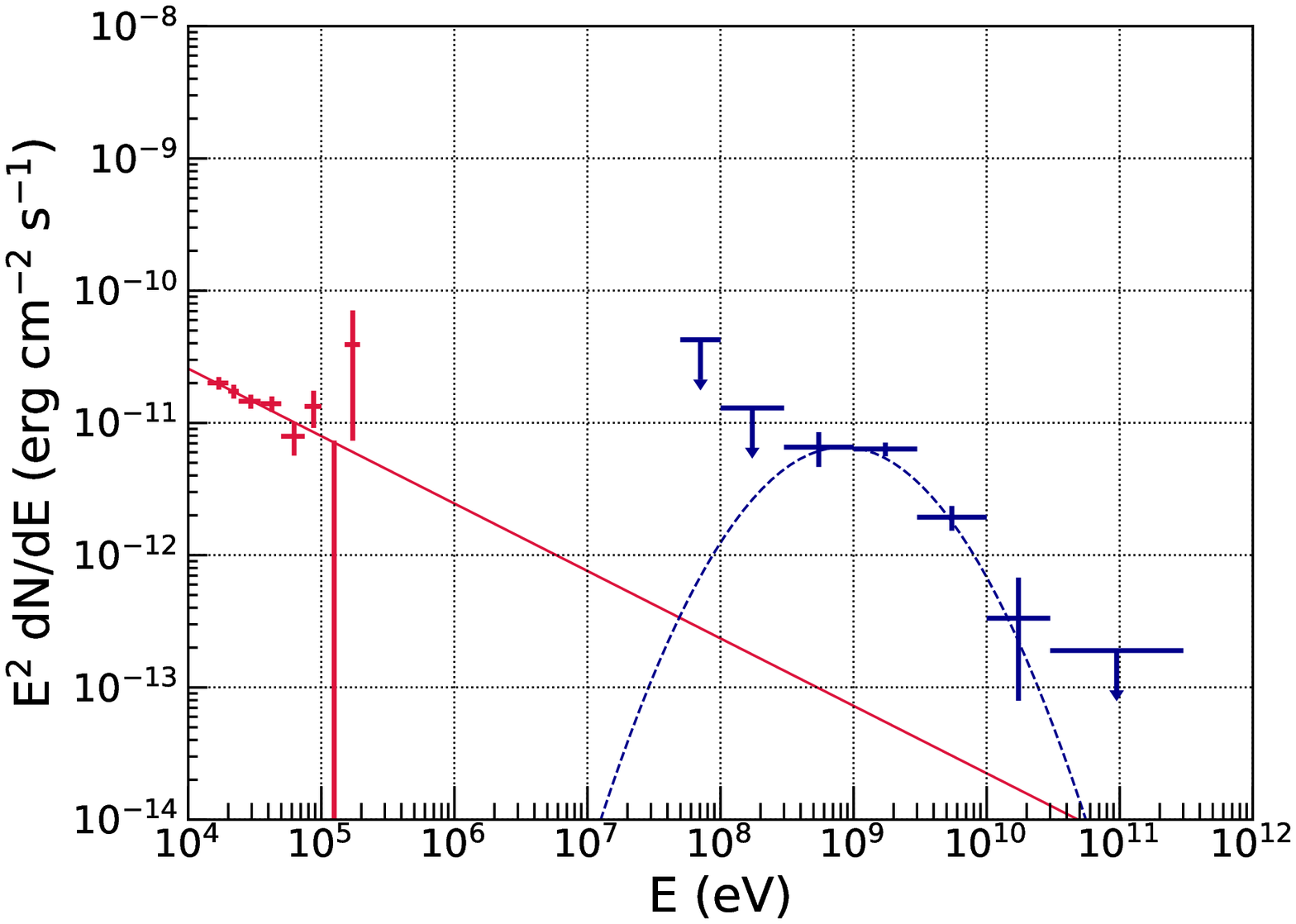}{0.32\textwidth}{XTE J1652$-$453 (No. 132)}
          }          
\gridline{\fig{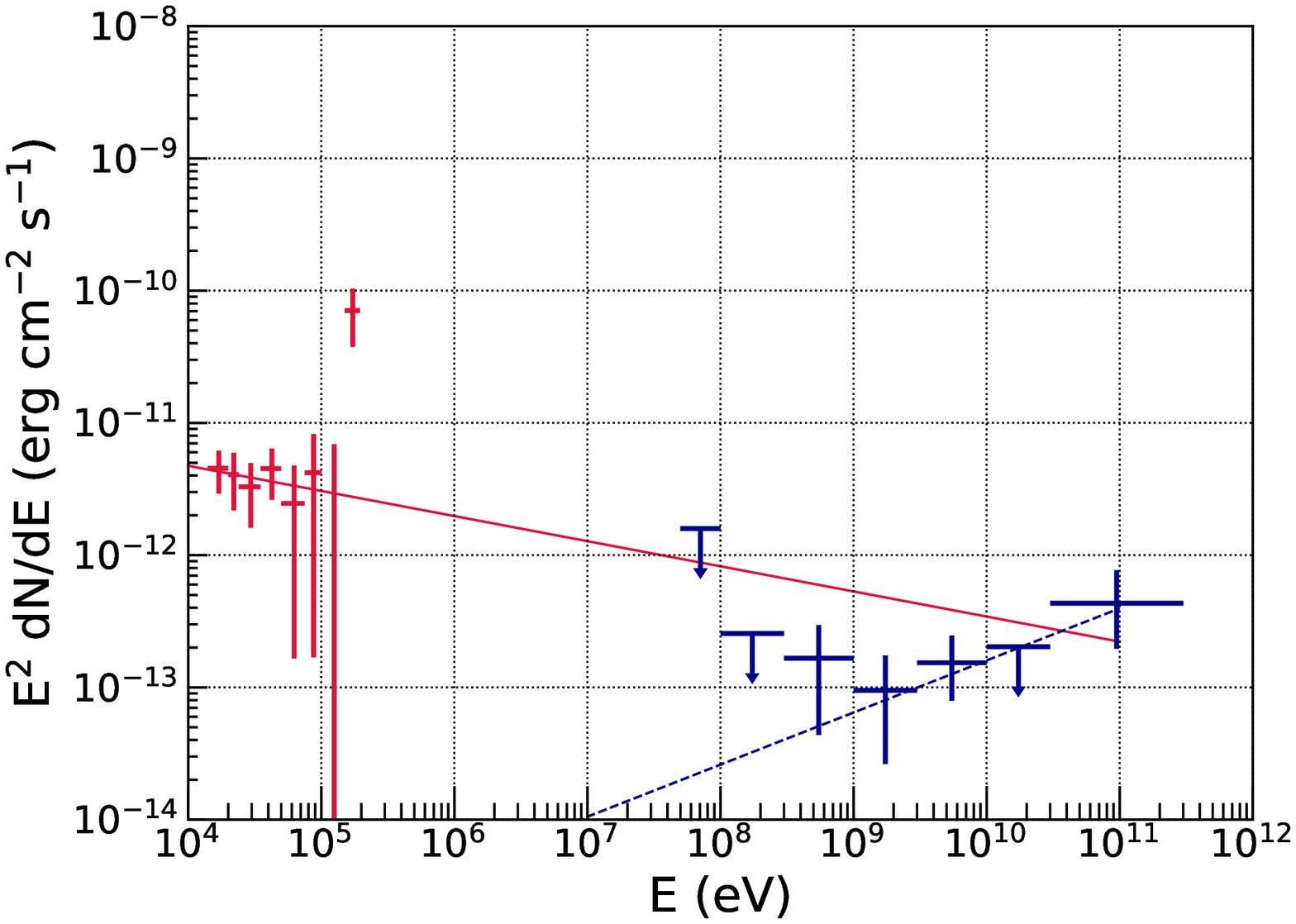}{0.32\textwidth}{CGCG 147$-$020 (No. 154)}
          \fig{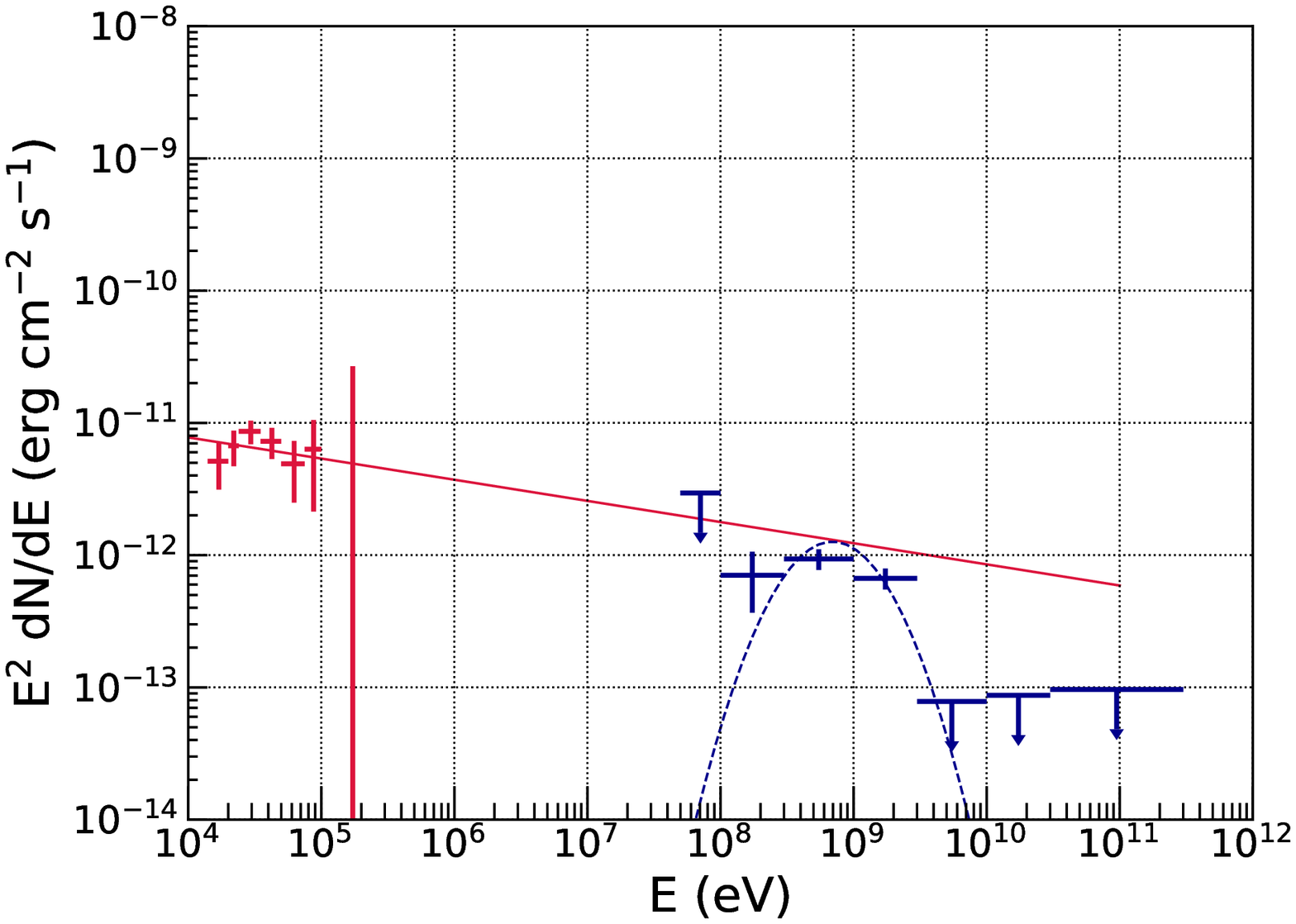}{0.32\textwidth}{2MASX J14080674$-$3023537 (No. 155)}
          \fig{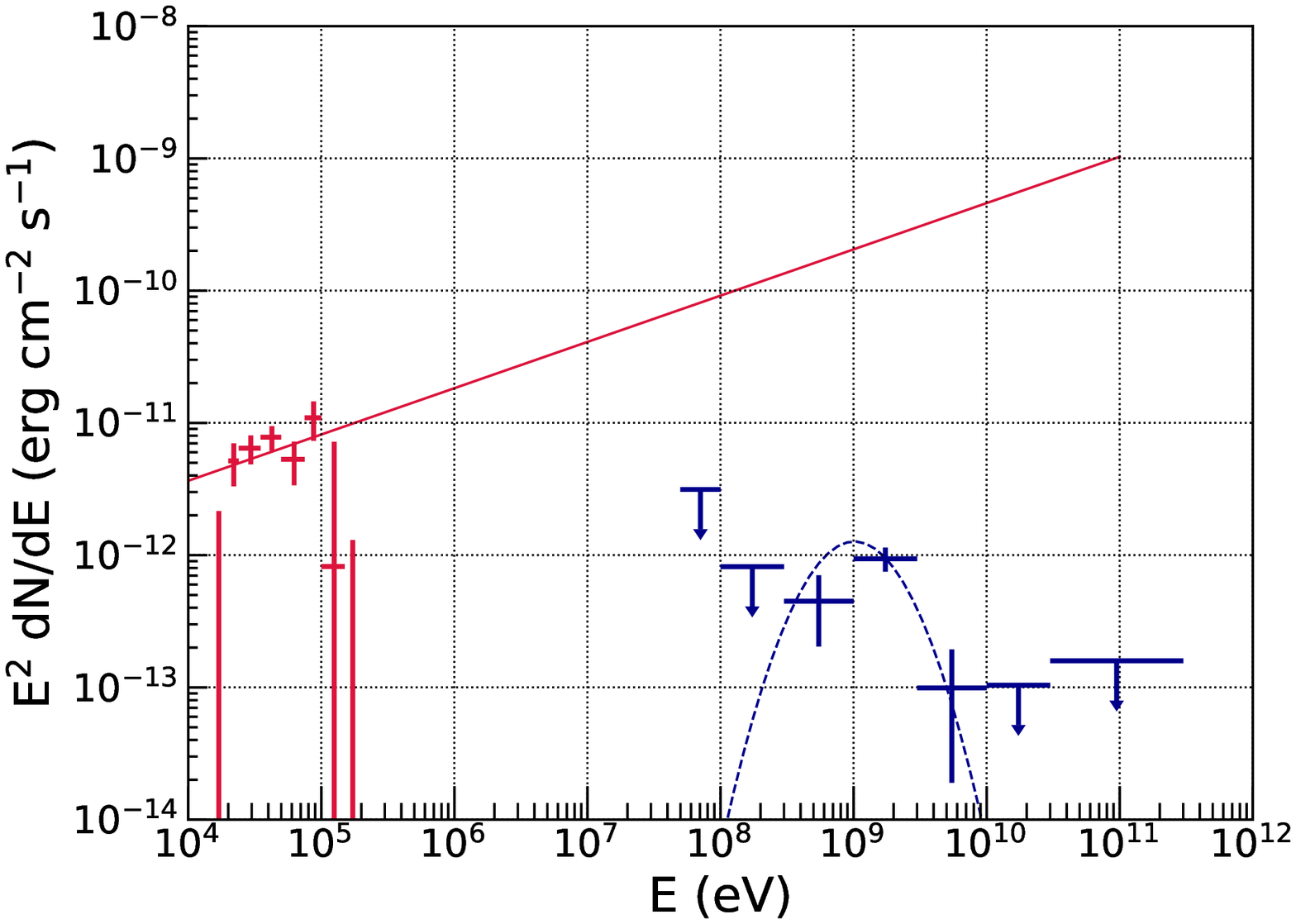}{0.32\textwidth}{XTE J1817$-$330 (No. 156)}
          }                 
\caption{
\acp{sed} of the unidentified point-like sources in the \bat\ (14--195 keV) and \lat\ (50 MeV--300 GeV) energy bands, shown in red and blue, respectively.
The red solid and blue dashed lines indicate the model spectrum taken from the \bat\ and \lat\ catalogs, respectively.
}
\label{fig:sed}
\end{center}
\end{figure*}

Here we report on the unidentified point-like sources found in our analysis and discuss possible associations.
The unidentified source is defined as the positionally matched source with its source type being unclear either in the \bat\ or \lat\ catalogs.
\figref{fig:sed} shows \acp{sed} of the 9 unidentified sources.
Each source is briefly described in the following.

\begin{enumerate}
\item No. 58 in \tabref{tab:point_source}: SWIFT J1254.9$+$1165 (U3\footnote{`U3' indicates unknown sources without soft X-ray counterparts in the \bat\ 105-month catalog.}) in the BAT catalog was matched with ON 187 (fsrq) in 4FGL-DR2. They are possibly associated, inferred from the FSRQ-like SED and the small separation of 0.006\degr.

\item No. 65: SWIFT J0949.1$+$4057 (confused source) in the BAT catalog was matched with 4C $+$40.24 (fsrq) in 4FGL-DR2. This association needs more investigation to be confirmed, particularly in the hard X-ray energy range that was uncertain due to the large errors. Deeper observations would give us a clue for such a faint source.

\item No. 75: PMN J0145-2733 (Unknown AGN) in the BAT catalog was matched with PKS 0142-278 (fsrq) in 4FGL-DR2. This could be likely an association, inferred from the FSRQ-like \ac{sed}. However, more X-ray observations would be necessary to precisely measure the upturn-like feature seen at $\sim 70$~keV in order to determine its origin and the association with the GeV gamma-ray radiation.

\item No. 130: GX 340$+$0 (LMXB) in the BAT catalog was matched with 4U 1642-45 (unk) in 4FGL-DR2. The association between these two sources is promising, since they have the same identification. The GeV emission with \lat, however, is unknown due to being located in a complex TeV gamma-ray emitting region, HESS J1648$-$458 \citep[see, e.g., ][]{HESS2012_westerlund1}. 
Beside the accreting neutron star 4U 1642$-$45, HESS J1648$-$458 contained PSR J1648$-$4611 and a star cluster Westerlund 1.
4U 1642$-$45 was unlikely responsible for the TeV gamma rays, inferred from the spatial extent and time variation. They argued that a single source scenario would favor the hadronic gamma-ray radiation produced by collisions of cosmic rays from Westerlund 1 with the \ac{ism}.

\item No. 131: SAX J1808.4$-$3658 (LMXB) in the BAT catalog was matched with SWIFT J1808.5$-$3655 (unknown) in 4FGL-DR2. Note that the counterpart of the \fermi\ source is not a firm association (i.e., SWIFT J1808.5$-$3655 was labeled with `ASSOC2'). This association ---the gamma-ray emission from the LMXB--- was previously reported and discussed in \cite{deOnaWilhelmi2015}.

\item No. 132: XTE J1652$-$453 (LMXB) in the BAT catalog was matched with SWIFT J1652.3$-$4520 (unknown) in 4FGL-DR2. Note that the counterpart of the \fermi\ source is not a firm association (i.e., SWIFT J1652.3$-$4520 was labeled with `ASSOC2'). They might be associated as the former case of SAX J1808.4$-$3658, although further investigation is needed to confirm the association.

\item No. 154--156: CGCG 147$-$020 (Sy2; No. 154), 2MASX J14080674$-$3023537 (Sy1.9; No. 155), and XTE J1817$-$330 (LMXB; No. 156) are the matched \bat\ sources, and they are unknown sources in 4FGL-DR2. 
These were faint, and thus the position uncertainty was large both in the BAT and LAT observations. The dedicated deeper observations are necessary for them to unveil the association and the nature.

\end{enumerate}

We conducted a time variation analysis of the unidentified point-like sources using 1-month lightcurves of the \bat\ catalog and 2-month lightcurves of 4FGL (the 8-yr \lat\ catalog).
No significant correlation between the hard X-ray and GeV gamma-ray radiation in the 2-month scale was found in any unidentified source, probably because of the poor statistics.
In the case of the binary system, timing analyses folded by the orbital period are necessary to track the variability correlation. This is beyond the scope of this paper and will be performed in the future publication.


\clearpage
\subsection{Unidentified extended sources} \label{sec:unID_extended}
\begin{figure*}[ht]
\begin{center}
\gridline{\fig{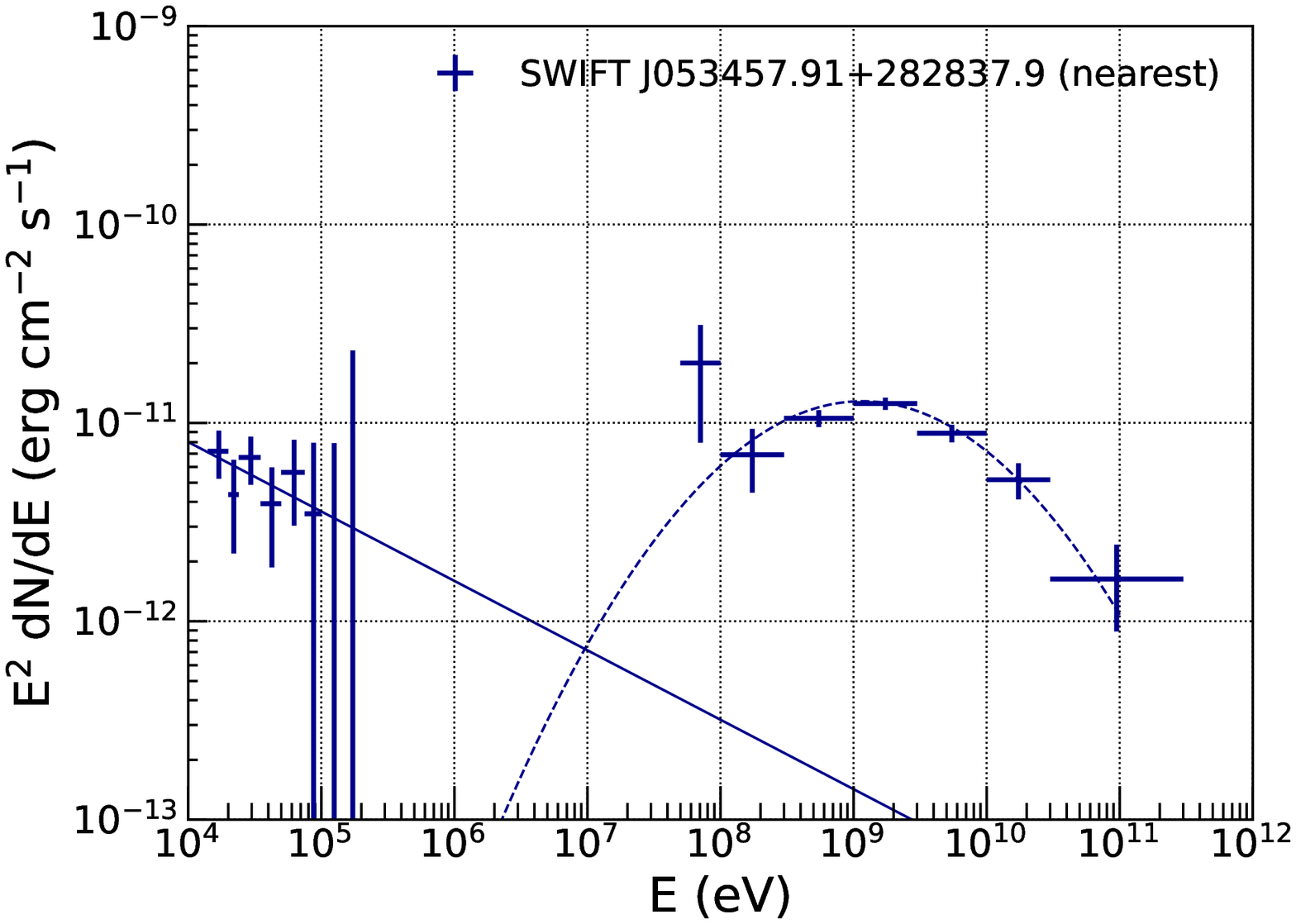}{0.32\textwidth}{Sim 147 (No. 17 in \tabref{tab:extended})}
          \fig{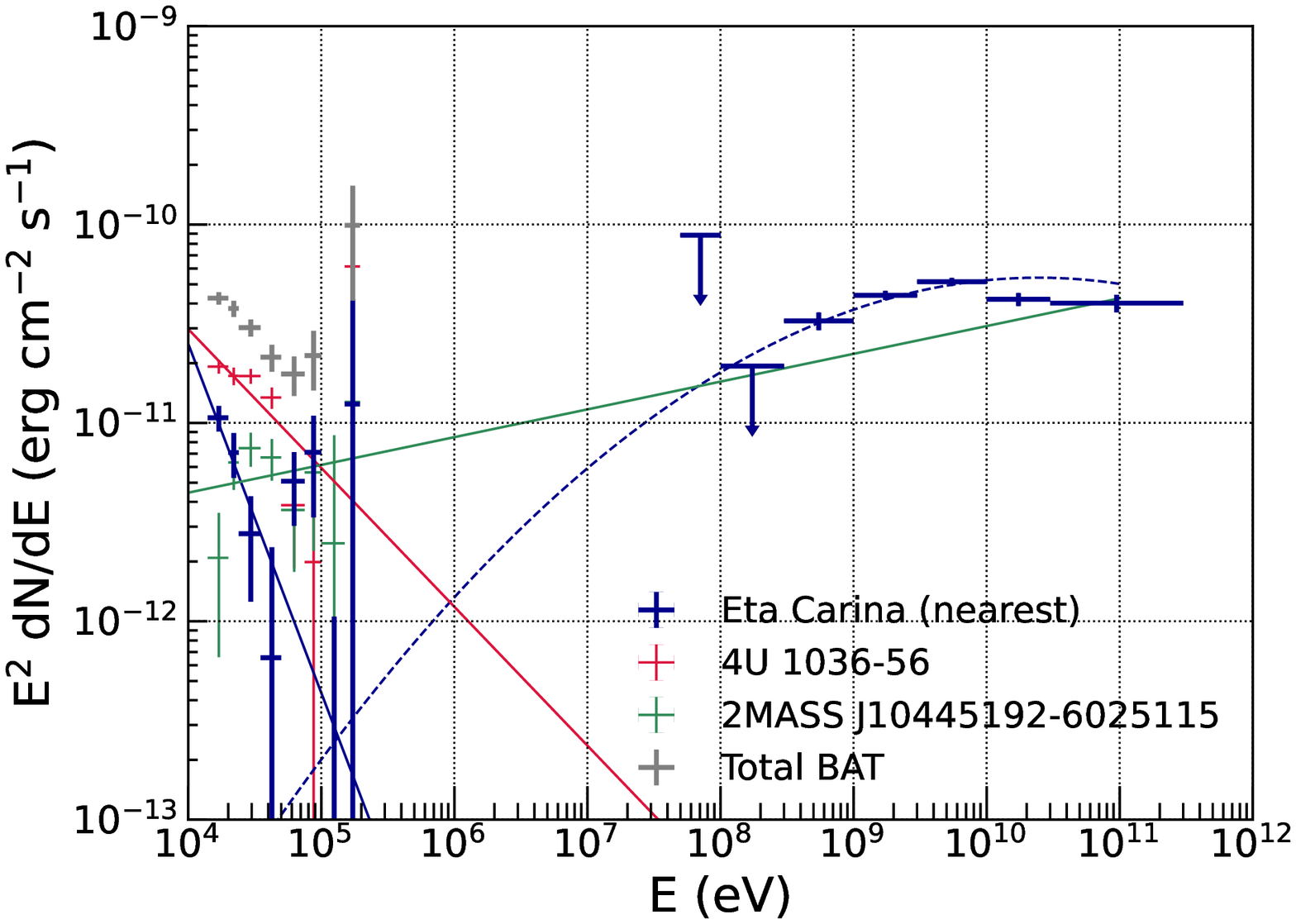}{0.32\textwidth}{FGES J1036.3$-$5833 (No. 29)}
}
\gridline{          
          \fig{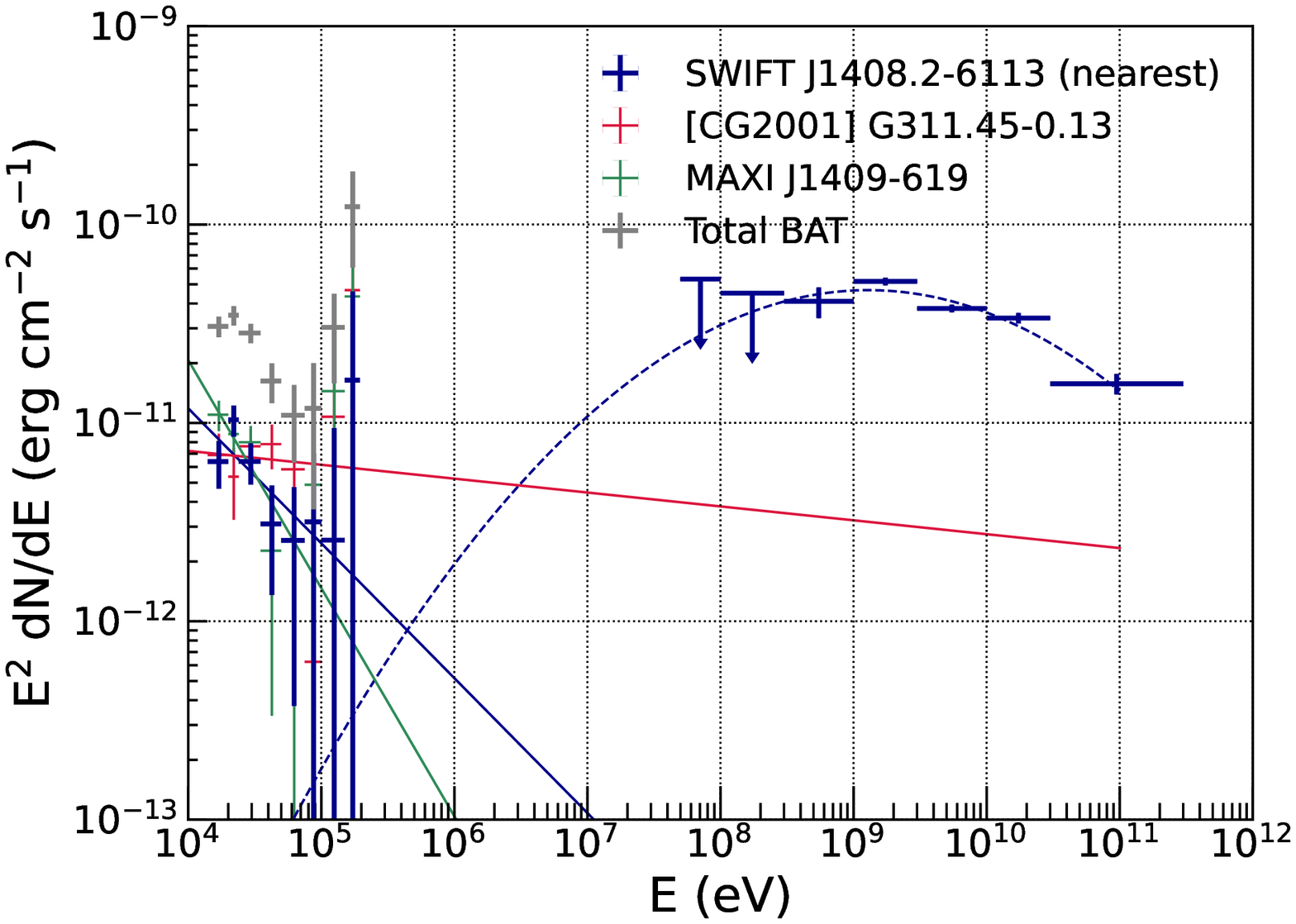}{0.32\textwidth}{FGES J1409.1$-$6121 (No. 30)}          
		  \fig{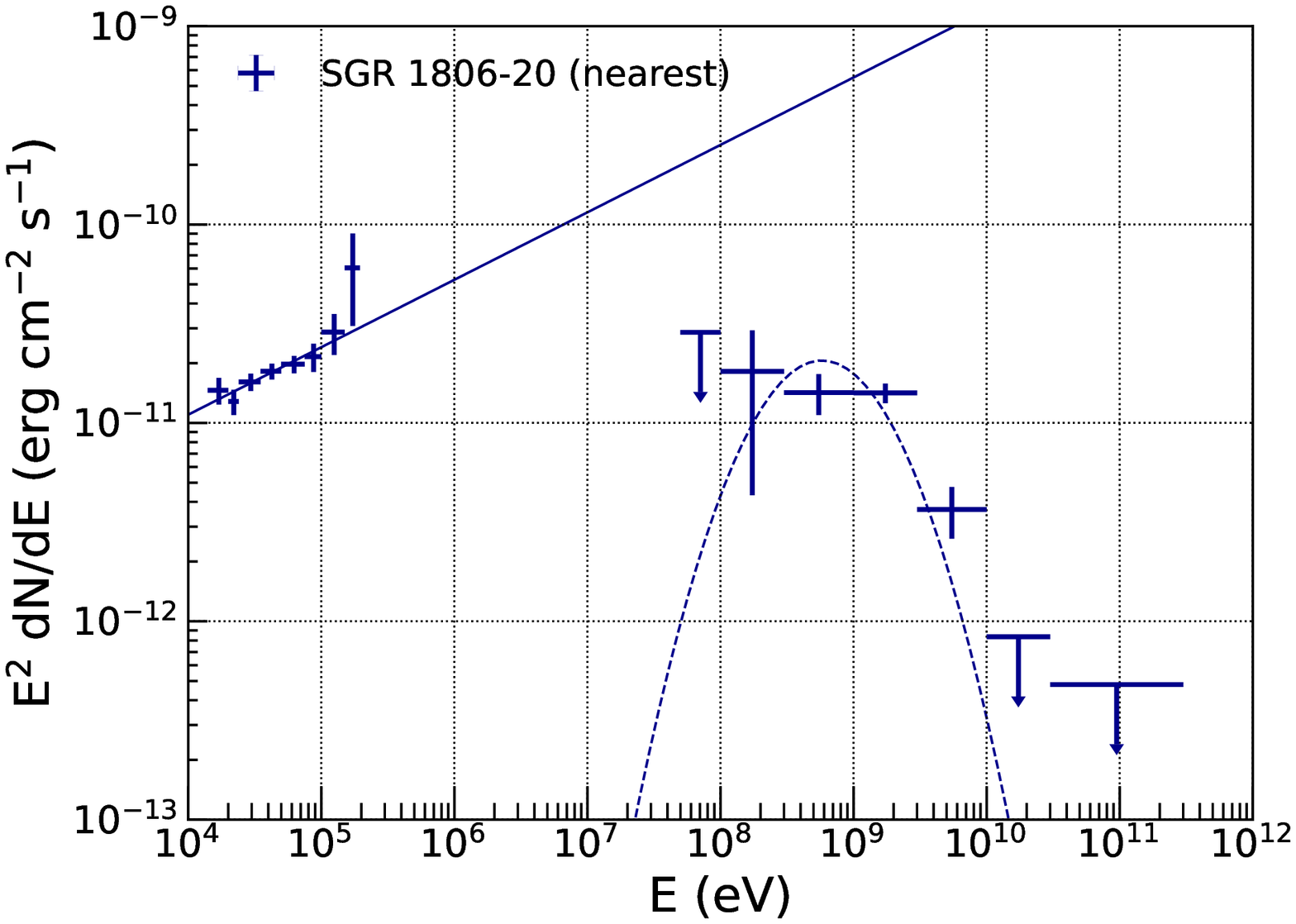}{0.32\textwidth}{HESS J1808$-$204 (No. 31)}
          }          
\caption{
\acp{sed} of unidentified extended sources.
The blue solid and dashed lines indicate the model spectra provided in the \bat\ and \lat\ catalogs, respectively.
For FGES J1036.3$-$5833 and FGES J1409.1$-$6121, \acp{sed} of all BAT counterparts and the total flux are also shown.
}
\label{fig:sed_extended}
\end{center}
\end{figure*}

We briefly describe the current status of the unidentified and extended sources in our study. 
Their \acp{sed} are illustrated in \figref{fig:sed_extended}.

\begin{enumerate}

\item No. 17 in \tabref{tab:extended}: Sim 147 (SNR) was spatially matched with SWIFT J053457.91$+$282837.9 (U2\footnote{U2 indicates a source of which its soft X-ray emission is detected from archival X-ray observation with S/N greater than 3.}) in the \bat\ catalog. Sim 147 is a middle-aged SNR, including a known PSR-PWN association inside its GeV gamma-ray extent of 1.5\degr\ \citep{Katsuta2012}. The matched source, SWIFT J053457.91$+$282837.9, was revealed to be a possible intermediate polar (i.e., a cataclysmic variable binary star system) by a periodic analysis of optical observations \citep{Halpern2018}. Therefore we suggest that these two sources are not associated and are false-matched. This would also be supported by the fact that the BAT source is located near the edge of the gamma-ray emission, and there exists the aforementioned PSR-PWN association close to the center of the SNR.

\item No. 29: FGES J1036.3$-$5833 (unidentified) hosts inside the extent Eta Carina (XRB), 4U 1036$-$56 (HMXB), and 2MASS J10445192$-$6025115 (star). This gamma-ray emission is largely extended with $\sim$2.5\degr\ in radius, and is remarkably variable in the 1-yr scale with the variability index of $\sim75$. The time variation could result from a variable source inside the gamma-ray extent (i.e., Eta Carina or 4U 1036$-$56). 

\item No. 30: FGES J1409.1$-$6121 (unidentified) has spatial coincidences with SWIFT J1408.2$-$6113 (CV), [CG2001] G311.45$-$0.13 (U2), and MAXI J1409$-$619 (Pulsar). The gamma-ray extent is $\sim$0.73\degr. The gamma-ray emission might be associated with [CG2001] G311.45$-$0.13, which could be a possible counterpart of a radio SNR G12.4$-$0.4 \citep{Doherty2003}. However, the hard spectrum in the \bat\ energy regime ($\Gamma \sim 2$) is not likely of origin of the X-ray radiation from the remnant. An alternative is MAXI J1409$-$619, a pulsar, which is located in the vicinity of SNR G12.4$-$0.4. Further investigation would be necessary to confirm the association.

\item No. 31: HESS J1808$-$204 (unidentified) was spatially matched with SGR 1806$-$20 (a pulsar, more like a magnetar) in the \bat\ catalog. \cite{Yeung2016} reported the possible association between the gamma-ray radiation with \lat\ and the magnetar, and later the origin (i.e., the gamma-ray emission powered by magnetic dissipation from SGR 1806$-$20) was discussed in \cite{HESS2018_HESSJ1808}. These studies, however, could not reach to a robust conclusion due to other plausible scenarios to account for the gamma-ray radiation.  

\end{enumerate}

\subsection{Future prospect}    \label{sec:future}

\begin{figure*}[ht!]
\begin{center}
\plotone{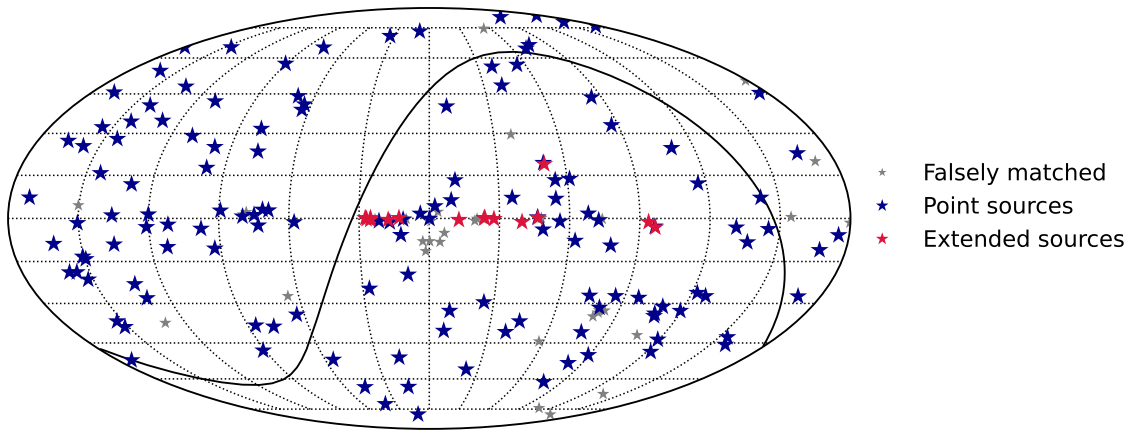}
\caption{
The cross-matched sources shown on the galactic coordinate.
The firmly matched point sources and extended sources are shown in blue and red, respectively, while the false matches are shown in grey.
The solid line indicates the declination of 0\degr.
}
\label{fig:sky}
\end{center}
\end{figure*}

Over 20 years ago, \comptel\ confirmed 25 steady MeV gamma-ray emitting sources based on the observational data with the flux sensitivity of $\sim 10 ^{-10}$~\flux\ \citep{Schonfelder2000}.
In the last decade, the sensitivity of the detectors in the neighboring energy bands (i.e., the hard X-ray and GeV gamma ray) has improved to  $< 10 ^{-11}$~\flux. 
This work reports 151 sources firmly matched between the latest \bat\ and \lat\ catalogs.
We present these cross-matched sources in the all-sky map in \figref{fig:sky}.
The matched catalog (\tabref{tab:point_source} and \tabref{tab:extended}) contains promising objects that are bright in the MeV energy range and are detectable with future instruments with a sensitivity being over one order of magnitude better than \comptel.
This catalog would be a helpful resource when devising a strategy for the ongoing projects of the MeV observation, such as {\it e-ASTROGAM} \citep{eASTROGAM2018}, {\it AMEGO} \citep{AMEGO2019}, {\it COSI} \citep{COSI_SMEX}, and {\it GRAMS} \citep{Aramaki2020}.
The cross-matched sources, combined with a simulation of diffuse emission, can be useful to predict the all sky image in the MeV energy channel. This will be presented in a future publication.


\clearpage
\section{Conclusions} \label{sec:conclusions}

We performed a cross-matched between the \bat\ 105-month catalog and the 4FGL-DR2 catalog.
We confirmed 
(1) 132 sources (115 firmly matched sources) by the spatial cross-match with the separation threshold of \rsep=0.08\degr,
(2) 31 sources (15 firmly matched sources) by the spatial cross-match for extended sources,
and (3) 24 sources (21 firmly matched sources) by the identification match.
The firmly matched sources (151 in total) predominantly consisted of blazars. Particularly, the proportion of FSRQs in the matched catalog was over twice as large as that of the 4FGL-DR2.
We found that most of \comptel\ sources were included in this study, and the cross-match with \ibis\ catalog could add 8 point-like and 4 extended sources.
Compared to the original catalogs, the distributions of physical parameters of the matched sources were characterized by the bimodal feature in the $\Gamma$ distribution, a higher flux, and larger variability index, resulting from the different source fractions.

\acknowledgments

We thank the anonymous referee for the fruitful comments and advise.
This work made use of data from the \swift\ and \fermi\ observatories.
N.T. acknowledges RIKEN iTHEMS Program.
H.Y. is supported by the Japan Society for the Promotion of Science (JSPS) KAKENHI grant No. 20K22355,
Y.I. is supported by JSPS KAKENHI grant Nos. JP16K13813, JP18H05458, and JP19K14772,
and H.O. is supported by JSPS KAKENHI grant Nos. 19H05185 and 19H01906.



%

\vspace{5mm}

\facilities{\bat, \lat}

\bibliography{MeV.bib} 

\bibliographystyle{aasjournal}




\movetabledown=2cm

\startlongtable
\begin{longrotatetable}
\begin{deluxetable*}{cccccccccccccccccc}
\tablecaption{Cross-matched point-like sources
\label{tab:point_source}
}
\tabletypesize{\footnotesize}
\tablehead{
\colhead{No.} & \colhead{\bat\ name} & \colhead{Type} & \colhead{\lat\ name} & \colhead{Type} & \colhead{Sep.} & \colhead{Flag} & \colhead{1FLE} & \colhead{BAT coord.} & \colhead{$\Gamma_{\rm BAT}$} & \colhead{$\Gamma_{\rm LAT}$} & \colhead{$F_{\rm BAT}$} & \colhead{$F_{\rm LAT}$} & \colhead{VarIndex} \\
\colhead{} &  \colhead{} & \colhead{ } & \colhead{ } & \colhead{ } & \colhead{(deg)}  & \colhead{} &  \colhead{} & \colhead{($\alpha_{\rm J200}$, $\delta_{\rm J2000}$)} &  \colhead{} & \colhead{} & \colhead{} & \colhead{} & 
} 
\startdata
1 & [HB89] 0537-441 & Beamed AGN & PKS 0537-441 & BLL & 0.063 & M & 1 & (84.63, -44.12) & 0.88 & 2.11 & 1.5 & 17 & 9.8e+03  \\ 
2 & [HB89] 0716+714 & Beamed AGN & S5 0716+71 & BLL & 0.01 & M & 1 & (110.5, 71.33) & 1.15 & 2.08 & 1.9 & 21 & 2.7e+03  \\ 
3 & Mrk 421 & Beamed AGN & Mkn 421 & BLL & 0.014 & M & 1 & (166.1, 38.21) & 2.76 & 1.78 & 14 & 42 & 1.3e+03  \\ 
4 & Mrk 501 & Beamed AGN & Mkn 501 & BLL & 0.001 & M & 0 & (253.5, 39.76) & 2.39 & 1.76 & 7.2 & 13 & 5.4e+02  \\ 
5 & BL Lac & Beamed AGN & BL Lac & BLL & 0.018 & M & 1 & (330.7, 42.27) & 1.76 & 2.2 & 3.5 & 23 & 3.5e+03  \\ 
6 & 1ES 0033+595 & Beamed AGN & 1ES 0033+595 & bll & 0.008 & M & 0 & (8.989, 59.84) & 2.81 & 1.76 & 2.6 & 4 & 2e+02  \\ 
7 & 2MASX J01155048+2515369 & Sy1 & RX J0115.7+2519 & bll & 0.018 & A & 0 & (18.97, 25.33) & 1.97 & 1.92 & 1.1 & 2 & 3.1e+02  \\ 
8 & SHBL J012308.7+342049 & Beamed AGN & 1ES 0120+340 & bll & 0.021 & M & 0 & (20.78, 34.37) & 2.94 & 1.7 & 1.1 & 0.3 & 15  \\ 
9 & B3 0133+388 & Beamed AGN & B3 0133+388 & bll & 0.052 & M & 0 & (24.13, 39.05) & 1.99 & 1.72 & 0.79 & 5.1 & 61  \\ 
10 & RBS 259 & Beamed AGN & 1RXS J015658.6-530208 & bll & 0.062 & M & 0 & (29.13, -53.04) & 2.31 & 1.74 & 0.73 & 0.59 & 45  \\ 
11 & 2MASX J02141794+5144520 & Beamed AGN & TXS 0210+515 & bll & 0.03 & M & 0 & (33.55, 51.77) & 2.58 & 1.9 & 1.4 & 0.53 & 13  \\ 
12 & QSO B0229+200 & Beamed AGN & 1ES 0229+200 & bll & 0.036 & M & 0 & (38.19, 20.29) & 2.28 & 1.79 & 2.3 & 0.37 & 6.8  \\ 
13 & ESO 416-G002 & Sy1.9 & PHL 1389 & bll & 0.068 & A & 0 & (38.83, -29.63) & 1.67 & 2.03 & 2.1 & 0.17 & 14  \\ 
14 & BZB J0244-5819 & Beamed AGN & RBS 0351 & bll & 0.031 & M & 0 & (41.19, -58.3) & 2.43 & 1.75 & 1 & 0.67 & 40  \\ 
15 & QSO B0347-121 & Beamed AGN & 1ES 0347-121 & bll & 0.018 & M & 0 & (57.37, -11.98) & 2.2 & 1.76 & 1.6 & 0.39 & 13  \\ 
16 & PKS 0352-686 & Beamed AGN & PKS 0352-686 & bll & 0.004 & M & 0 & (58.28, -68.53) & 2.52 & 1.67 & 1.2 & 0.33 & 8.3  \\ 
17 & PKS 0426-380 & Beamed AGN & PKS 0426-380 & bll & 0.059 & M & 0 & (67.14, -37.89) & 2.56 & 2.1 & 0.38 & 21 & 3.5e+03  \\ 
18 & PKS 0548-322 & Beamed AGN & PKS 0548-322 & bll & 0.054 & M & 0 & (87.69, -32.27) & 3.23 & 1.89 & 1.8 & 0.39 & 9  \\ 
19 & PMN J0640-1253 & Beamed AGN & TXS 0637-128 & bll & 0.056 & M & 0 & (100.1, -12.87) & 2.56 & 1.65 & 0.79 & 0.56 & 4.9  \\ 
20 & 2MASX J07103005+5908202 & Beamed AGN & 1H 0658+595 & bll & 0.008 & M & 0 & (107.6, 59.14) & 2.28 & 1.69 & 2.4 & 0.67 & 27  \\ 
21 & 2MASS J09303759+4950256 & Beamed AGN & 1ES 0927+500 & bll & 0.069 & M & 0 & (142.5, 49.88) & 2.59 & 1.82 & 0.74 & 0.23 & 16  \\ 
22 & 2MASS J09343014-1721215 & Beamed AGN & RXC J0934.4-1721 & bll & 0.039 & M & 0 & (143.6, -17.34) & 2.73 & 1.94 & 0.79 & 0.21 & 8  \\ 
23 & 2MASX J10311847+5053358 & Beamed AGN & 1ES 1028+511 & bll & 0.02 & M & 0 & (157.9, 50.9) & 2.85 & 1.74 & 0.78 & 1.2 & 13  \\ 
24 & 2MASX J11033765-2329307 & Beamed AGN & 1ES 1101-232 & bll & 0.045 & M & 0 & (165.9, -23.47) & 2.53 & 1.77 & 1.1 & 0.55 & 13  \\ 
25 & 2MASX J11363009+6737042 & Beamed AGN & RX J1136.5+6737 & bll & 0.033 & M & 0 & (174.1, 67.64) & 2.33 & 1.75 & 1.3 & 0.63 & 28  \\ 
26 & FBQS J1221+3010 & Beamed AGN & PG 1218+304 & bll & 0.008 & M & 0 & (185.3, 30.16) & 2.94 & 1.71 & 1.1 & 4.4 & 60  \\ 
27 & [HB89] 1415+259 & Beamed AGN & 1E 1415.6+2557 & bll & 0.041 & M & 0 & (214.4, 25.72) & 2.61 & 1.45 & 0.59 & 0.33 & 8.1  \\ 
28 & 1ES 1426+428 & Beamed AGN & H 1426+428 & bll & 0.027 & M & 0 & (217.1, 42.66) & 2.56 & 1.63 & 2.1 & 0.88 & 16  \\ 
29 & [HB89] 1803+784 & Beamed AGN & S5 1803+784 & bll & 0.057 & M & 1 & (269.9, 78.47) & 1.93 & 2.21 & 0.91 & 5.4 & 2.8e+02  \\ 
30 & QSO B1959+650 & Beamed AGN & 1ES 1959+650 & bll & 0.019 & M & 0 & (300, 65.16) & 2.67 & 1.82 & 2.9 & 11 & 1.5e+03  \\ 
31 & 2MASX J23470479+5142179 & Beamed AGN & 1ES 2344+514 & bll & 0.005 & M & 0 & (356.8, 51.69) & 2.66 & 1.81 & 1 & 3.2 & 60  \\ 
32 & H 2356-309 & Beamed AGN & H 2356-309 & bll & 0.058 & M & 0 & (359.8, -30.58) & 2.28 & 1.82 & 1.5 & 0.53 & 12  \\ 
33 & 3C 454.3 & Beamed AGN & 3C 454.3 & FSRQ & 0.011 & M & 1 & (343.5, 16.15) & 1.5 & 2.4 & 16 & 1e+02 & 5.6e+04  \\ 
34 & [HB89] 2230+114 & Beamed AGN & CTA 102 & FSRQ & 0.031 & M & 1 & (338.2, 11.71) & 1.49 & 2.29 & 3 & 49 & 7.5e+04  \\ 
35 & PKS 2227-088 & Beamed AGN & PKS 2227-08 & FSRQ & 0.072 & M & 1 & (337.5, -8.492) & 1.46 & 2.59 & 1.7 & 4.2 & 6.4e+02  \\ 
36 & [HB89] 2142-758 & Beamed AGN & PKS 2142-75 & FSRQ & 0.074 & M & 1 & (327.1, -75.58) & 1.41 & 2.44 & 1.5 & 5.6 & 2.8e+03  \\ 
37 & QSO B2013+370 & Beamed AGN & MG2 J201534+3710 & FSRQ & 0.036 & M & 0 & (303.9, 37.21) & 2.13 & 2.45 & 1.8 & 7.5 & 1.3e+02  \\ 
38 & PKS 1830-21 & Beamed AGN & PKS 1830-211 & FSRQ & 0.017 & M & 1 & (278.4, -21.07) & 1.47 & 2.53 & 8.7 & 19 & 2.5e+03  \\ 
39 & 3C 345 & Beamed AGN & 3C 345 & FSRQ & 0.026 & M & 0 & (250.8, 39.81) & 1.17 & 2.4 & 2.1 & 3 & 2e+02  \\ 
40 & PKS 1622-29 & Beamed AGN & PKS B1622-297 & FSRQ & 0.034 & M & 1 & (246.6, -29.86) & 1.32 & 2.56 & 1.6 & 3.9 & 4.3e+02  \\ 
41 & PKS 1510-08 & Beamed AGN & PKS 1510-089 & FSRQ & 0.025 & M & 1 & (228.2, -9.081) & 1.32 & 2.38 & 6.7 & 42 & 5.9e+03  \\ 
42 & 3C 279 & Beamed AGN & 3C 279 & FSRQ & 0.015 & M & 1 & (194.1, -5.799) & 1.32 & 2.29 & 3.9 & 45 & 3e+04  \\ 
43 & 3C 273 & Beamed AGN & 3C 273 & FSRQ & 0.009 & M & 1 & (187.3, 2.047) & 1.75 & 2.7 & 42 & 11 & 7e+03  \\ 
44 & PG 1222+216 & Beamed AGN & 4C +21.35 & FSRQ & 0.021 & M & 1 & (186.2, 21.4) & 1.7 & 2.34 & 2.5 & 20 & 2e+04  \\ 
45 & 4C +49.22 & Beamed AGN & 4C +49.22 & FSRQ & 0.017 & M & 1 & (178.3, 49.5) & 1.83 & 2.41 & 1.3 & 1.6 & 1.2e+03  \\ 
46 & [HB89] 0836+710 & Beamed AGN & 4C +71.07 & FSRQ & 0.004 & M & 1 & (130.3, 70.89) & 1.7 & 2.82 & 7 & 4.8 & 3.4e+03  \\ 
47 & PMN J0641-0320 & Beamed AGN & PMN J0641-0320 & FSRQ & 0.046 & M & 0 & (100.5, -3.362) & 0.96 & 2.68 & 2.2 & 2.5 & 3.6e+02  \\ 
48 & PKS 0528+134 & Beamed AGN & PKS 0528+134 & FSRQ & 0.032 & M & 0 & (82.74, 13.57) & 1.25 & 2.56 & 1.8 & 2.2 & 4.6e+02  \\ 
49 & PKS 0402-362 & Beamed AGN & PKS 0402-362 & FSRQ & 0.017 & M & 1 & (60.97, -36.07) & 1.91 & 2.53 & 1.1 & 7.4 & 6.1e+03  \\ 
50 & PKS 2325+093 & Beamed AGN & PKS 2325+093 & fsrq & 0.012 & M & 1 & (351.9, 9.663) & 1.4 & 2.69 & 3 & 1.8 & 3.7e+02  \\ 
51 & 87GB 215950.2+503417 & Beamed AGN & NRAO 676 & fsrq & 0.016 & M & 1 & (330.4, 50.82) & 1.78 & 2.66 & 1.9 & 5.1 & 3.1e+03  \\ 
52 & PKS 2149-306 & Beamed AGN & PKS 2149-306 & fsrq & 0.015 & M & 1 & (328, -30.46) & 1.61 & 2.85 & 8.9 & 3 & 9.6e+02  \\ 
53 & [HB89] 1921-293 & Beamed AGN & PKS B1921-293 & fsrq & 0.063 & M & 0 & (291.2, -29.18) & 2.04 & 2.39 & 1.6 & 2.3 & 2.8e+02  \\ 
54 & 2MASS J16561677-3302127 & Beamed AGN & 2MASS J16561677-3302127 & fsrq & 0.024 & M & 0 & (254.1, -33.04) & 1.55 & 2.79 & 6.2 & 1.5 & 1.3e+02  \\ 
55 & [HB89] 1354+195 & Beamed AGN & 4C +19.44 & fsrq & 0.079 & M & 0 & (209.3, 19.29) & 2.02 & 2.76 & 0.87 & 0.4 & 39  \\ 
56 & [HB89] 1334-127 & Beamed AGN & PKS 1335-127 & fsrq & 0.013 & M & 0 & (204.4, -12.95) & 2.19 & 2.42 & 1.3 & 1.8 & 95  \\ 
57 & PKS 1329-049 & Beamed AGN & PKS 1329-049 & fsrq & 0.032 & M & 1 & (203, -5.153) & 1.51 & 2.51 & 1.5 & 3.5 & 2.2e+03  \\ 
58 & SWIFT J1254.9+1165 & U3 & ON 187 & fsrq & 0.006 & U & 0 & (193.7, 11.65) & 1.72 & 2.79 & 1.3 & 0.52 & 28  \\ 
59 & 4C +04.42 & Beamed AGN & 4C +04.42 & fsrq & 0.047 & M & 0 & (185.6, 4.219) & 1.45 & 2.79 & 3.6 & 1.9 & 1.6e+02  \\ 
60 & FBQS J1159+2914 & Beamed AGN & Ton 599 & fsrq & 0.045 & M & 1 & (179.9, 29.23) & 1.84 & 2.19 & 0.75 & 12 & 7.7e+03  \\ 
61 & 7C 1150+3324 & Beamed AGN & B2 1150+33A & fsrq & 0.053 & M & 0 & (178.2, 33.09) & 1.83 & 3.01 & 1 & 0.28 & 19  \\ 
62 & PKS 1143-696 & Beamed AGN & PKS 1143-696 & fsrq & 0.077 & M & 0 & (176.5, -69.9) & 1.69 & 2.69 & 1.5 & 0.56 & 35  \\ 
63 & PKS 1127-14 & Beamed AGN & PKS 1127-14 & fsrq & 0.065 & M & 0 & (172.5, -14.8) & 1.88 & 2.69 & 2.9 & 0.89 & 4.1e+02  \\ 
64 & [HB89] 1039+811 & Beamed AGN & S5 1039+81 & fsrq & 0.022 & M & 0 & (161.2, 80.92) & 1.67 & 2.78 & 1.2 & 0.92 & 47  \\ 
65 & SWIFT J0949.1+4057 & confused source & 4C +40.24 & fsrq & 0.063 & U & 0 & (147.3, 40.57) & 2.46 & 2.61 & 0.73 & 0.31 & 32  \\ 
66 & CGRaBS J0805+6144 & Beamed AGN & TXS 0800+618 & fsrq & 0.048 & M & 0 & (121.3, 61.75) & 1.35 & 2.8 & 1.8 & 0.67 & 2.2e+02  \\ 
67 & B2 0743+25 & Beamed AGN & B2 0743+25 & fsrq & 0.042 & M & 0 & (116.6, 25.81) & 1.43 & 2.86 & 3.6 & 0.67 & 88  \\ 
68 & PKS 0637-752 & Beamed AGN & PKS 0637-75 & fsrq & 0.038 & M & 0 & (99.02, -75.28) & 2.0 & 2.7 & 1.7 & 1.3 & 4.5e+02  \\ 
69 & [HB89] 0552+398 & Beamed AGN & B2 0552+39A & fsrq & 0.02 & M & 0 & (88.9, 39.81) & 1.54 & 2.76 & 2.1 & 1.8 & 1.5e+02  \\ 
70 & [HB89] 0537-286 & Beamed AGN & PKS 0537-286 & fsrq & 0.045 & M & 0 & (84.97, -28.7) & 1.33 & 2.73 & 2.9 & 1.6 & 1.1e+02  \\ 
71 & PKS 0524-460 & Beamed AGN & PKS 0524-460 & fsrq & 0.035 & M & 0 & (81.32, -46.01) & 1.37 & 2.38 & 1.6 & 0.35 & 7.3  \\ 
72 & [HB89] 0403-132 & Beamed AGN & PKS 0403-13 & fsrq & 0.061 & M & 0 & (61.36, -13.14) & 1.78 & 2.55 & 1.1 & 0.87 & 80  \\ 
73 & 4C +50.11 & Beamed AGN & NRAO 150 & fsrq & 0.017 & M & 0 & (59.9, 50.97) & 1.51 & 2.66 & 1.7 & 3.3 & 7.7e+02  \\ 
74 & [HB89] 0212+735 & Beamed AGN & S5 0212+73 & fsrq & 0.073 & M & 1 & (34.38, 73.81) & 1.55 & 2.94 & 3.5 & 1.4 & 41  \\ 
75 & PMN J0145-2733 & Unknown AGN & PKS 0142-278 & fsrq & 0.042 & U & 1 & (26.21, -27.54) & 1.43 & 2.6 & 1.1 & 0.98 & 8.1e+02  \\ 
76 & ESO 354- G 004 & Sy1.9 & PMN J0151-3605 & bcu & 0.007 & A & 0 & (27.87, -36.13) & 1.98 & 2.28 & 1.1 & 0.22 & 8  \\ 
77 & 4C +33.06 & Beamed AGN & 4C +33.06 & bcu & 0.049 & M & 0 & (46.19, 33.8) & 1.86 & 2.57 & 1.2 & 0.3 & 33  \\ 
78 & PKS 0706-15 & Beamed AGN & PKS 0706-15 & bcu & 0.024 & M & 0 & (107.3, -15.44) & 3.42 & 1.81 & 0.74 & 0.46 & 19  \\ 
79 & PKS 0723-008 & Beamed AGN & PKS 0723-008 & bcu & 0.03 & M & 0 & (111.5, -0.942) & 1.75 & 2.06 & 1.5 & 1.1 & 22  \\ 
80 & 2MASX J07332681+5153560 & Beamed AGN & NVSS J073326+515355 & bcu & 0.057 & M & 0 & (113.4, 51.93) & 2.32 & 1.81 & 0.82 & 0.28 & 12  \\ 
81 & IGR J13109-5552 & Sy1 & PMN J1310-5552 & bcu & 0.03 & D & 0 & (197.7, -55.91) & 1.56 & 2.82 & 2.5 & 0.72 & 78  \\ 
82 & PMN J1508-4953 & Beamed AGN & PMN J1508-4953 & bcu & 0.013 & M & 0 & (227.2, -49.87) & 1.15 & 2.84 & 2.9 & 1.9 & 36  \\ 
83 & SGR A* & Galactic Center & Galactic Centre & bcu & 0.016 & D & 0 & (266.4, -29.01) & 2.69 & 2.34 & 11 & 42 & 2.6  \\ 
84 & PKS 1936-623 & Beamed AGN & PKS 1936-623 & bcu & 0.017 & M & 1 & (295.3, -62.16) & 1.32 & 2.43 & 1.8 & 3.6 & 1.2e+03  \\ 
85 & SWIFT J1943536.21+211822.9 & Beamed AGN & MG2 J194359+2118 & bcu & 0.028 & M & 0 & (296, 21.31) & 2.11 & 1.51 & 2.8 & 0.74 & 15  \\ 
86 & B2 2023+33 & Beamed AGN & B2 2023+33 & bcu & 0.008 & M & 0 & (306.3, 33.68) & 1.49 & 2.63 & 1.3 & 2.9 & 1.2e+02  \\ 
87 & RX J2056.6+4940 & Beamed AGN & RGB J2056+496 & bcu & 0.007 & M & 0 & (314.2, 49.66) & 2.62 & 1.86 & 1.3 & 2.1 & 25  \\ 
88 & RBS 1895 & Beamed AGN & RBS 1895 & bcu & 0.014 & M & 0 & (341.7, -52.12) & 2.51 & 1.71 & 0.7 & 0.2 & 19  \\ 
89 & 1RXS J225146.9-320614 & Beamed AGN & 1RXS J225146.9-320614 & bcu & 0.044 & M & 0 & (342.9, -32.1) & 2.01 & 1.79 & 1.4 & 0.15 & 11  \\ 
90 & PKS 0521-36 & Beamed AGN & PKS 0521-36 & AGN & 0.018 & M & 1 & (80.76, -36.46) & 1.92 & 2.46 & 3.5 & 5.5 & 6.8e+02  \\ 
91 & Cen A & Beamed AGN & Cen A & RDG & 0.012 & M & 1 & (201.4, -43.02) & 1.88 & 2.64 & 1.3e+02 & 6.3 & 18  \\ 
92 & 3C 120 & Beamed AGN & 3C 120 & RDG & 0.039 & M & 0 & (68.3, 5.356) & 2.01 & 2.74 & 9.5 & 1.5 & 2.8e+02  \\ 
93 & NGC 1275 & Beamed AGN & NGC 1275 & RDG & 0.009 & M & 1 & (49.95, 41.51) & 3.82 & 2.11 & 8.3 & 33 & 4.8e+03  \\ 
94 & PKS 2300-18 & Beamed AGN & PKS 2300-18 & rdg & 0.069 & M & 0 & (345.8, -18.7) & 2.03 & 2.18 & 1.5 & 0.25 & 9.9  \\ 
95 & 3C 303 & Sy1 & 3C 303 & rdg & 0.046 & D & 0 & (220.7, 52.06) & 2.39 & 2.05 & 0.57 & 0.17 & 8.5  \\ 
96 & PICTOR A & Sy2 & Pictor A & rdg & 0.045 & D & 0 & (79.95, -45.77) & 2.05 & 2.43 & 3.7 & 0.41 & 7.3  \\ 
97 & QSO B0309+411 & Beamed AGN & B3 0309+411B & rdg & 0.039 & M & 0 & (48.26, 41.29) & 1.58 & 2.69 & 1.5 & 0.41 & 25  \\ 
98 & NGC 4945 & Sy2 & NGC 4945 & sbg & 0.003 & D & 0 & (196.4, -49.47) & 1.5 & 2.27 & 28 & 1.1 & 10  \\ 
99 & M 82 & Starburst galaxy & M 82 & sbg & 0.029 & M & 0 & (148.9, 69.64) & 3.39 & 2.22 & 0.48 & 1.1 & 6.1  \\ 
100 & NGC 1068 & Sy1.9 & NGC 1068 & sbg & 0.004 & D & 0 & (40.66, -0.004) & 1.82 & 2.33 & 3.8 & 0.65 & 18  \\ 
101 & Circinus Galaxy & Sy2 & Circinus galaxy & sey & 0.01 & M & 0 & (213.3, -65.34) & 2.09 & 2.26 & 27 & 0.67 & 14  \\ 
102 & 1H 0323+342 & Beamed AGN & 1H 0323+342 & nlsy1 & 0.046 & M & 0 & (51.21, 34.17) & 1.62 & 2.82 & 2.7 & 1.9 & 2e+02  \\ 
103 & 3C 380 & Beamed AGN & 3C 380 & css & 0.028 & M & 1 & (277.4, 48.74) & 1.52 & 2.42 & 1.5 & 1.8 & 69  \\ 
104 & Cas A & SNR & Cas A & snr & 0.008 & M & 0 & (350.8, 58.82) & 3.33 & 1.97 & 6.5 & 6.2 & 4.5  \\ 
105 & Tycho SNR & SNR & Tycho & snr & 0.01 & M & 0 & (6.326, 64.14) & 3.03 & 2.18 & 1.3 & 0.98 & 4.2  \\ 
106 & SNR G068.8+02.6 & SNR & PSR J1952+3252 & PSR & 0.016 & D & 0 & (298.3, 32.89) & 2.27 & 2.28 & 0.92 & 15 & 6  \\ 
107 & SNR G21.5-00.9 & SNR & PSR J1833-1034 & PSR & 0.01 & D & 0 & (278.4, -10.57) & 2.26 & 2.5 & 7.4 & 8.2 & 3  \\ 
108 & 4U 1820-30 & LMXB & PSR J1823-3021A & PSR & 0.025 & F & 0 & (275.9, -30.36) & 5.2 & 2.21 & 95 & 1.4 & 2.6  \\ 
109 & SLX 1744-299 & LMXB & PSR J1747-2958 & PSR & 0.047 & F & 0 & (266.9, -30) & 3.25 & 2.56 & 13 & 16 & 15  \\ 
110 & 1RXS J122758.8-485343 & CV & PSR J1227-4853 & PSR & 0.007 & D & 0 & (187, -48.89) & 1.85 & 2.38 & 3.7 & 2.3 & 74  \\ 
111 & PSR J1124-5916 & Pulsar & PSR J1124-5916 & PSR & 0.054 & M & 0 & (171.1, -59.31) & 2.47 & 2.46 & 0.83 & 6.1 & 8.6  \\ 
112 & Vela Pulsar & Pulsar & PSR J0835-4510 & PSR & 0.006 & M & 1 & (128.8, -45.18) & 1.97 & 2.23 & 18 & 9.4e+02 & 4.6  \\ 
113 & PSR B0540-69 & Pulsar & PSR J0540-6919 & PSR & 0.027 & M & 0 & (85.02, -69.35) & 1.93 & 2.47 & 4.9 & 2.8 & 8.1  \\ 
114 & 2MASX J04372814-4711298 & Sy1 & PSR J0437-4715 & PSR & 0.079 & F & 0 & (69.41, -47.21) & 1.96 & 2.35 & 0.99 & 1.7 & 9.6  \\ 
115 & PSR J1811-1925 & Pulsar & PSR J1811-1925 & psr & 0.014 & M & 0 & (272.9, -19.42) & 2.07 & 2.14 & 3.5 & 1.1 & 8.6  \\ 
116 & Crab & Pulsar & Crab Nebula$^\dagger$ & PWN & 0.003 & M & 0 & (83.63, 22.02) & 2.17 & 3.8 & 2.3e+03 & 16 & 6.1e+02  \\ 
117 & IGR J17461-2853 & molecular cloud & PWN G0.13-0.11 & pwn & 0.079 & F & 0 & (266.5, -28.89) & 1.7 & 2.46 & 3.3 & 7 & 7.4  \\ 
118 & Cyg X-3 & HMXB & Cyg X-3 & HMB & 0.072 & M & 0 & (308.1, 40.96) & 3.0 & 2.66 & 2.5e+02 & 3.5 & 83  \\ 
119 & RX J1826.2-1450 & HMXB & LS 5039 & HMB & 0.002 & M & 0 & (276.6, -14.85) & 1.62 & 2.61 & 3.3 & 27 & 9  \\ 
120 & 2XMM J130247.6-635008 & HMXB & PSR B1259-63 & HMB & 0.069 & M & 0 & (195.6, -63.85) & 1.2 & 2.75 & 2 & 1.6 & 1.9e+02  \\ 
121 & LS I +61 303 & HMXB & LSI +61 303 & HMB & 0.014 & M & 1 & (40.16, 61.24) & 1.73 & 2.38 & 3.2 & 47 & 2.5e+02  \\ 
122 & Cyg X-1 & HMXB & Cyg X-1 & hmb & 0.035 & M & 0 & (299.6, 35.2) & 1.9 & 2.14 & 1.7e+03 & 0.65 & 12  \\ 
123 & V395 Car & LMXB & 2S 0921-630 & lmb & 0.053 & M & 0 & (140.5, -63.31) & 5.38 & 2.23 & 0.56 & 0.19 & 2.7  \\ 
124 & Eta Carina & XRB & Eta Carinae & BIN & 0.005 & M & 0 & (161.3, -59.68) & 3.76 & 2.31 & 0.78 & 19 & 47  \\ 
125 & 4U 2129+12 & LMXB & NGC 7078 & glc & 0.026 & F & 0 & (322.5, 12.17) & 2.66 & 2.62 & 7.7 & 0.42 & 23  \\ 
126 & XB 1832-330 & LMXB & NGC 6652 & glc & 0.015 & F & 0 & (278.9, -32.99) & 2.26 & 2.35 & 18 & 0.48 & 13  \\ 
127 & 4U 1746-37 & LMXB & NGC 6441 & glc & 0.021 & F & 0 & (267.6, -37.05) & 5.45 & 2.4 & 8.3 & 1.5 & 8.2  \\ 
128 & ESO 520-27 & GC & Terzan 5 & glc & 0.009 & M & 0 & (267, -24.78) & 4.32 & 2.37 & 2.6 & 7.9 & 5.3  \\ 
129 & 4U 1722-30 & LMXB & Terzan 2 & glc & 0.049 & F & 0 & (261.9, -30.8) & 2.51 & 2.37 & 43 & 0.67 & 7.4  \\ 
130 & GX 340+0 & LMXB & 4U 1642-45 & unk & 0.058 & U & 0 & (251.4, -45.61) & 5.59 & 2.61 & 86 & 4.5 & 5.4  \\ 
131 & SAX J1808.4-3658 & LMXB & (SWIFT J1808.5-3655) & unk & 0.043 & U & 0 & (272.1, -36.99) & 2.34 & 2.44 & 3.7 & 0.44 & 12  \\ 
132 & XTE J1652-453 & LMXB & (SWIFT J1652.3-4520) & unk & 0.061 & U & 0 & (253.1, -45.34) & 2.51 & 2.58 & 3.1 & 1.9 & 9.2  \\ 
133 & PKS 2005-489 & Beamed AGN & PKS 2005-489 & BLL & 0.088 & M & 0 & (302.5, -48.87) & 2.42 & 1.83 & 0.6 & 3.2 & 1.5e+02  \\ 
134 & 87GB 050246.4+673341 & Beamed AGN & 1ES 0502+675 & bll & 0.094 & M & 0 & (76.92, 67.53) & 2.5 & 1.58 & 0.9 & 2.4 & 56  \\ 
135 & 2MASX J03252346-5635443 & Beamed AGN & 1RXS J032521.8-563543 & bll & 0.082 & M & 0 & (51.47, -56.53) & 2.06 & 1.95 & 0.85 & 0.58 & 36  \\ 
136 & PKS 0607-549 & Beamed AGN & PKS 0607-549 & bcu & 0.137 & M & 0 & (92.21, -55.08) & 2.19 & 2.68 & 0.62 & 0.68 & 77  \\ 
137 & B2 0920+39 & Beamed AGN & B2 0920+39 & bcu & 0.107 & M & 0 & (140.8, 38.78) & 1.44 & 2.69 & 1.2 & 0.59 & 91  \\ 
138 & 8C 1849+670 & Beamed AGN & S4 1849+67 & FSRQ & 0.087 & M & 0 & (282.1, 67.05) & 2.72 & 2.29 & 0.64 & 4.4 & 1.9e+03  \\ 
139 & RBS 0315 & Beamed AGN & TXS 0222+185 & fsrq & 0.089 & M & 0 & (36.26, 18.8) & 1.73 & 2.95 & 3.1 & 0.5 & 52  \\ 
140 & 4C +32.14 & Beamed AGN & NRAO 140 & fsrq & 0.121 & M & 0 & (54.12, 32.29) & 1.67 & 2.8 & 4.4 & 1.3 & 46  \\ 
141 & PKS 2008-159 & Beamed AGN & PKS 2008-159 & fsrq & 0.083 & M & 0 & (302.8, -15.75) & 2.41 & 2.82 & 1.3 & 0.48 & 18  \\ 
142 & PKS 2052-47 & Beamed AGN & PKS 2052-47 & fsrq & 0.183 & M & 1 & (313.8, -47.16) & 2.19 & 2.45 & 1.8 & 6 & 1.3e+03  \\ 
143 & PKS 2145+06 & Beamed AGN & PKS 2145+06 & fsrq & 0.142 & M & 0 & (327, 6.936) & 1.9 & 2.8 & 1.7 & 0.54 & 39  \\ 
144 & [HB89] 0834-201 & Beamed AGN & PKS 0834-20 & fsrq & 0.197 & M & 0 & (129.2, -20.25) & 1.43 & 2.91 & 1.4 & 0.58 & 25  \\ 
145 & 1RXS J174036.3+521155 & Beamed AGN & 4C +51.37 & fsrq & 0.222 & M & 1 & (265.2, 51.97) & 1.9 & 2.47 & 0.87 & 2.1 & 8e+02  \\ 
146 & 2MASX J06230765-6436211 & Beamed AGN & RX J062308.0-643619 & fsrq & 0.115 & M & 0 & (95.85, -64.61) & 1.98 & 3.04 & 1.2 & 0.25 & 7.9  \\ 
147 & 87GB 162418.8+435342 & Beamed AGN & MG4 J162551+4346 & fsrq & 0.249 & M & 0 & (246.5, 43.81) & 2.04 & 2.91 & 1.2 & 0.23 & 14  \\ 
148 & PKS 2331-240 & Beamed AGN & PKS 2331-240 & agn & 0.262 & M & 0 & (353.5, -23.69) & 1.4 & 2.5 & 1.6 & 0.44 & 20  \\ 
149 & PKS 2153-69 & Sy2 & PKS 2153-69 & rdg & 0.125 & D & 0 & (329.4, -69.7) & 1.59 & 2.87 & 1.5 & 0.37 & 5.2  \\ 
150 & 3C 111.0 & Sy1.2 & 3C 111 & rdg & 0.1 & D & 1 & (64.59, 38.02) & 2.0 & 2.74 & 12 & 1.5 & 40  \\ 
151 & 3C 309.1 & Beamed AGN & 3C 309.1 & css & 0.098 & M & 0 & (224.5, 71.72) & 1.8 & 2.5 & 0.77 & 0.43 & 2.1e+02  \\ 
152 & PSR J1420-6048 & Pulsar & PSR J1420-6048 & PSR & 0.272 & M & 0 & (215.3, -60.58) & 2.24 & 2.42 & 1 & 14 & 14  \\ 
153 & PSR J1723-2837 & Pulsar & PSR J1723-2837 & psr & 0.361 & M & 0 & (260.8, -28.64) & 0.88 & 2.58 & 1.9 & 0.58 & 12  \\ 
154 & CGCG 147-020 & Sy2 & (SWIFT J0725.8+3000) & unk & 0.282 & U & 0 & (111.4, 30.02) & 2.19 & 1.61 & 0.92 & 0.094 & 15  \\ 
155 & 2MASX J14080674-3023537 & Sy1.9 & (SWIFT J1408.1-3024) & unk & 0.15 & U & 0 & (212.1, -30.38) & 2.16 & 2.59 & 1.6 & 0.24 & 16  \\ 
156 & XTE J1817-330 & LMXB & (SWIFT J1817.8-3301) & unk & 0.144 & U & 0 & (274.3, -32.98) & 1.65 & 2.5 & 1.8 & 0.22 & 8.8  \\ 
\enddata
\tablecomments{ 
Flux is in units of $10^{-11}$~\flux.
In 4FGL-DR2, source types with capital letters (e.g., BLL) indicate firm associations, while those with small letters (e.g., bll) are associations.
The \lat\ name in parenthesis indicate the identifications of ASSOC2 (less firm associations).
No. 133--156 are matched by their names, not spatially matched.
\\
$^\dagger$ Crab in 4FGL-DR2 has three entries, PSR J0534+2200, Crab nebula (synchrotron radiation), and Crab nebula (\ac{ic} scattering). Only the synchrotron component of the nebula is shown here.
}
\end{deluxetable*}
\end{longrotatetable}

\startlongtable
\begin{longrotatetable}	
\begin{deluxetable*}{ccc cccc | ccc cc | c }
\tablecaption{Cross-matched extended sources
\label{tab:extended}
}
\tablehead{
\colhead{No.} & \colhead{\lat\ name} & \colhead{Type} & \colhead{LAT coord.} & \colhead{Ext.} & \colhead{$\Gamma_{\rm LAT}$} & \colhead{$F_{\rm LAT}$} &  \colhead{\bat\ name} & \colhead{Type} & \colhead{Sep.} & \colhead{$\Gamma_{\rm BAT}$} & \colhead{$F_{\rm BAT}$}  &  \colhead{Flag} \\
\colhead{} &  \colhead{} & \colhead{ } & \colhead{($\alpha_{\rm J200}$, $\delta_{\rm J2000}$)} & \colhead{(deg)}  & \colhead{} & \colhead{} & \colhead{} &\colhead{} &\colhead{(deg)} & \colhead{} &\colhead{}  & \colhead{}
} 
\startdata
1 & LMC & GAL & (80, -68.75) & 3.0 & 2.19 & 11 & 2MASX J05052442-6734358 & Unknown AGN & 1.7 & 2.02 & 1 & F  \\  
 & &  & & & & & SWIFT J045106.8-694803 & HMXB & 2.8  & 2.5 & 3.3  \\  
 & &  & & & & & IGR J05007-7047 & HMXB & 2.6  & 2.4 & 2.2  \\  
 & &  & & & & & LMC X-4 & HMXB & 2.6  & 2.8 & 32.7  \\  
 & &  & & & & & RX J0531.2-6609 & HMXB & 2.7  & 2.8 & 1.35  \\  
 & &  & & & & & LMC X-1 & HMXB & 2.0  & 2.6 & 3.21  \\  
 & &  & & & & & PSR B0540-69 & Pulsar & 1.9  & 1.9 & 4.93  \\  
 & &  & & & & & XMMU J054134.7-682550 & HMXB & 2.0  & 2.9 & 2.42  \\  
 & &  & & & & & [RSG2010] A & HMXB & 1.9  & 2.8 & 0.566  \\  
\hline
2 & LMC-FarWest & ... & (75.25, -69.75) & 0.9 & 2.1 & 2 & SWIFT J045106.8-694803 & HMXB & 0.8 & 2.48 & 3.3 & F  \\  
\hline
3 & LMC-30DorWest & ... & (82.5, -69) & 0.9 & 2.12 & 4.4 & RSG2010 A & HMXB & 1.0 & 2.82 & 0.57 & F  \\  
\hline
4 & LMC-North & ... & (82.97, -66.65) & 0.6 & 2.0 & 2.3 & LMC X-4 & HMXB & 0.3 & 2.83 & 33 & F  \\  
 & &  & & & & & RX J0531.2-6609 & HMXB & 0.4  & 2.8 & 1.35  \\  
\hline
5 & SMC & GAL & (14.5, -72.75) & 1.5 & 2.2 & 2.5 & IGR J01054-7253 & HMXB & 0.3 & 3.46 & 0.34 & F  \\  
 & &  & & & & & RX J0052.1-7319 & HMXB & 0.7  & 2.7 & 1.58  \\  
 & &  & & & & & RX J0053.8-7226 & HMXB & 0.5  & 2.5 & 1.06  \\  
 & &  & & & & & XTE J0103-728 & HMXB & 0.4  & 3.2 & 1.34  \\  
 & &  & & & & & SXP 202 & HMXB & 0.4  & 2.9 & 0.153  \\  
\hline
6 & Cen A Lobes & RDG & (201, -43.5) & 2.5 & 2.51 & 5.2 & Cen A & Beamed AGN & 0.5 & 1.88 & 1.4e+02 & M  \\  
\hline
7 & Vela X & PWN & (128.3, -45.19) & 0.9 & 2.18 & 13 & Vela Pulsar & Pulsar & 0.4 & 1.97 & 18 & M  \\  
 & &  & & & & & SWIFT J0837.8-4440 & U2 & 1.0  & 2.5 & 1.16  \\  
\hline
8 & HESS J1420-607 & PWN & (215.1, -60.78) & 0.1 & 2.0 & 3.7 & Rabbit & Pulsar & 0.2 & 1.53 & 0.8 & M  \\  
\hline
9 & MSH 15-52 & PWN & (228.6, -59.16) & 0.2 & 1.83 & 5.3 & PSR B1509-58 & Pulsar & 0.0 & 1.85 & 26 & M  \\  
\hline
10 & HESS J1616-508 & PWN & (244.1, -50.91) & 0.3 & 2.05 & 12 & PSR J1617-5055 & Pulsar & 0.2 & 2.05 & 1.5 & M  \\  
\hline
11 & HESS J1825-137 & PWN & (276.1, -13.85) & 0.8 & 1.75 & 14 & IGR J18246-1425 & Pulsar & 0.6 & 2.8 & 1.9 & M  \\  
 & &  & & & & & XMMSL1 J182155.0-134719 & HMXB & 0.6  & 2.7 & 1.52  \\  
\hline
12 & HESS J1841-055 & PWN & (280.2, -5.55) & 0.6 & 1.98 & 13 & AX J1841.0-0535 & HMXB & 0.1 & 1.91 & 2.5 & M  \\  
 & &  & & & & & 1E 1841-045 & Pulsar & 0.6  & 1.3 & 10.7  \\  
\hline
13 & HESS J1632-478 & pwn & (247.9, -47.94) & 0.3 & 1.76 & 3 & AX J1631.9-4752 & Pulsar & 0.1 & 2.84 & 31 & M  \\  
\hline
14 & HESS J1837-069 & pwn & (279.1, -6.866) & 0.5 & 2.04 & 22 & PSR J1838-0655 & Pulsar & 0.3 & 1.71 & 6.9 & M  \\  
\hline
15 & HESS J1837-069 & pwn & (279.7, -7.067) & 0.5 & 1.85 & 7 & PSR J1838-0655 & Pulsar & 0.3 & 1.71 & 6.9 & M  \\  
\hline
16 & SNR G150.3+04.5 & SNR & (66.82, 55.55) & 1.5 & 1.68 & 4.1 & XTE J0421+560 & HMXB & 1.2 & 2.27 & 1.2 & F  \\  
\hline
17 & Sim 147 & SNR & (85.1, 27.94) & 1.5 & 2.18 & 5.7 & SWIFT J053457.91+282837.9 & U2 & 1.3 & 2.35 & 1.2 & U  \\  
\hline
18 & Monoceros & SNR & (99.86, 6.93) & 3.5 & 2.3 & 9.1 & 2MASX J06262702+0727287 & Unknown AGN & 3.2 & 1.87 & 1.6 & F  \\  
\hline
19 & RX J0852.0-4622 & SNR & (133, -46.34) & 1.0 & 1.79 & 12 & PSR J0855-4644 & Pulsar & 0.8 & 2.06 & 1 & M  \\  
\hline
20 & SNR G337.0-00.1 & SNR & (249.1, -47.52) & 0.1 & 2.34 & 10 & SGR 1627-41 & Gamma-ray source & 0.1 & 1.81 & 1.4 & A  \\  
 & &  & & & & & IGR J16358-4726 & Pulsar & 0.1  & 2.1 & 0.39  \\  
\hline
21 & gamma Cygni & SNR & (305.3, 40.52) & 0.6 & 1.96 & 10 & 2MASX J20183871+4041003 & Sy2 & 0.5 & 2.03 & 2.6 & F  \\  
\hline
22 & RX J1713.7-3946 & SNR & (258.4, -39.76) & 0.6 & 1.71 & 7 & SWIFT J1712.9-4002 & U1 & 0.3 & 3.25 & 1.3 & M  \\  
 & &  & & & & & SNR G347.3-0.5 & SNR & 0.3  & 3.1 & 1.99  \\  
\hline
23 & Cygnus X & SFR & (307.2, 41.17) & 3.0 & 2.09 & 1.2e+02 & Cyg X-3 & HMXB & 0.7 & 3.0 & 2.5e+02 & F  \\  
 & &  & & & & & 2MASX J20183871+4041003 & Sy2 & 1.9  & 2.0 & 2.62  \\  
 & &  & & & & & SSTSL2 J203705.58+415005.3 & Beamed AGN & 1.7  & 5.2 & 1.36  \\  
\hline
24 & HESS J1632-478 & spp & (248.3, -47.77) & 0.6 & 2.17 & 25 & AX J1631.9-4752 & Pulsar & 0.2 & 2.84 & 31 & M  \\  
 & &  & & & & & 4U 1630-47 & LMXB & 0.4  & 2.7 & 30.3  \\  
 & &  & & & & & IGR J16328-4726 & HMXB & 0.4  & 3.1 & 2.38  \\  
 & &  & & & & & SGR 1627-41 & Gamma-ray source & 0.5  & 1.8 & 1.38  \\  
 & &  & & & & & IGR J16358-4726 & Pulsar & 0.6  & 2.1 & 0.39  \\  
\hline
25 & HESS J1809-193 & spp & (272.6, -19.43) & 0.5 & 2.36 & 5.1 & PSR J1811-1925 & Pulsar & 0.3 & 2.07 & 3.5 & A  \\  
 & &  & & & & & XTE J1810-189 & LMXB & 0.4  & 2.2 & 8.21  \\  
\hline
26 & HESS J1813-178 & spp & (273.3, -17.62) & 0.6 & 2.34 & 15 & IGR J18135-1751 & SNR & 0.2 & 1.92 & 4.1 & M  \\  
 & &  & & & & & GX 13+1 & LMXB & 0.6  & 5.7 & 37.9  \\  
\hline
27 & W 41 & spp & (278.6, -8.78) & 0.2 & 2.13 & 11 & Swift J1834.9-0846 & star & 0.2 & 2.13 & 0.96 & F  \\  
\hline
28 & Kes 73 & spp & (280.2, -4.89) & 0.3 & 2.37 & 7.2 & 1E 1841-045 & Pulsar & 0.1 & 1.33 & 11 & M  \\  
\hline
29 & (FGES J1036.3-5833) & ... & (159.1, -58.56) & 2.5 & 1.93 & 29 & Eta Carina & XRB & 1.6 & 3.76 & 0.78 & U  \\  
 & &  & & & & & 4U 1036-56 & HMXB & 1.8  & 2.7 & 2.82  \\  
 & &  & & & & & 2MASS J10445192-6025115 & star & 2.2  & 1.9 & 1.49  \\  
\hline
30 & (FGES J1409.1-6121) & ... & (212.3, -61.35) & 0.7 & 2.16 & 25 & SWIFT J1408.2-6113 & CV & 0.2 & 2.68 & 1.1 & U  \\  
 & &  & & & & & [CG2001] G311.45-0.13 & U2 & 0.7  & 2.1 & 1.7  \\  
 & &  & & & & & MAXI J1409-619 & Pulsar & 0.6  & 3.1 & 1.17  \\  
\hline
31 & (HESS J1808-204) & ... & (272, -20.48) & 0.6 & 2.57 & 4.9 & SGR 1806-20 & Pulsar & 0.1 & 1.66 & 5.3 & U  \\ 
\hline
\enddata
\tablecomments{ 
$^\dagger$\ In 4FGL, HESS J1837$-$069 and HESS J1632$-$478 are extended and have two entries.
\\
Flux is in units of $10^{-11}$~\flux.
The nearest BAT source is listed at the top for the source matched with more than one BAT sources.
The LAT extent of a major axis is shown here.
}
\end{deluxetable*}
\end{longrotatetable}



\begin{longrotatetable}
\begin{deluxetable*}{ccc | cc | cc | l p{4cm} } 
\tablecaption{Cross-match with \comptel\ sources
\label{tab:comptel}
}
\tabletypesize{\footnotesize}
\tablehead{
\colhead{No.} & \colhead{\comptel } & \colhead{Type} & \colhead{\bat} & \colhead{Type}  & \colhead{\lat} & \colhead{Type} & \colhead{No. in \tabref{tab:point_source}}
& \colhead{Note}
\\
} 
\startdata
1 & PSR B1951+32 & PSR & SNR G068.8+02.6 & SNR & PSR J1952+3252               & PSR   & 106  \\ 
2 & PSR B0531+21 & PSR & Crab & Pulsar & PSR J0534+2200               & PSR   & 114  \\ 
3 & PSR J0633+1746 & PSR & ... & ... & PSR J0633+1746               & PSR   & ...  & Matched with only \fermi. Faint in the hard X-ray. \\ 
4 & PSR B0656+14 & PSR & ... & ... & PSR J0659+1414               & PSR   & ...  & Matched with only \fermi. Faint in the hard X-ray. \\ 
5 & PSR B0833-45 & PSR & Vela Pulsar & Pulsar & PSR J0835-4510               & PSR   & 112  \\ 
6 & PSR B1055-52 & PSR & ... & ... & PSR J1057-5226               & PSR   & ...  & Matched with only \fermi. Faint in the hard X-ray. \\ 
7 & PSR B1509-58 & PSR & PSR B1509-58 & Pulsar & MSH 15-52                    & PWN   & 10 in \tabref{tab:extended}  \\ 
8 & GRO J1823-12 & Galactic & RX J1826.2-1450 & HMXB & LS5039 & HMB & 120  \\ 
9 & Cygnus X-1 & Galactic & Cyg X-1 & HMXB & Cyg X-1                      & hmb   & 123  \\ 
10 & GRO J2227+61 & Galactic & ... & ... & PSR J2229+6114               & PSR   & ... & It has a BAT source (SWIFT J2221.6$+$5952) at 1.7\degr. \\ 
11 & GT 0236+610 & Galactic & LS I +61 303 & HMXB & LSI +61 303                  & HMB   & 122  \\ 
12 & Nova Per 1992 & Galactic & ... & ... & ... & ... & ...  & No match. X-ray transient. \\ 
13 & Crab Unpulsed & Galactic & Crab & Pulsar & PSR J0534+2200               & PSR   & 114  \\ 
14 & Vela/Carina & Galactic$^\dagger$ & ... & ... & 4FGL J0853.9-5501 & unk & ...  \\ 
15 & 3C 273 & AGN & 3C 273 & Beamed AGN & 3C 273                       & FSRQ  & 43  \\ 
16 & 3C 279 & AGN & 3C 279 & Beamed AGN & 3C 279                       & FSRQ  & 42  \\ 
17 & 3C 454.3 & AGN & 3C 454.3 & Beamed AGN & 3C 454.3                     & FSRQ  & 33  \\ 
18 & CTA 102 & AGN & [HB89] 2230+114 & Beamed AGN & CTA 102                      & FSRQ  & 34  \\ 
19 & Centaurus A & AGN & Cen A & Beamed AGN & Cen A                        & RDG   & 91  \\ 
20 & GRO J0516-609 & AGN & ... & ... & ... & ... & ...  & Unknown flaring source. It has a \fermi\ source (PMN J0507$-$6104) at 1.03\degr. \\ 
21 & GRO J1224+2155 & AGN & PG 1222+216 & Beamed AGN & 4C +21.35                    & FSRQ  & 44  \\ 
22 & PKS 0208-512 & AGN & ... & ... & PKS 0208-512                 & FSRQ  & ... & Matched with only \fermi.   \\ 
23 & PKS 0528+134 & AGN & PKS 0528+134 & Beamed AGN & PKS 0528+134                 & FSRQ  & 48  \\ 
24 & PKS 1622-297 & AGN & PKS 1622-29 & Beamed AGN & PKS B1622-297                & FSRQ  & 40  \\ 
25 & GRO J 1753+57  & Unknown$^\dagger$ & ... & ... & ... & ... & ... & No match.  \\ 
26 & GRO J 1040+48  & Unknown & ... & ... & ... & ... & ...  & No match.  \\ 
27 & GRO J 1214+06  & Unknown & 2MASX J12150077+0500512 & Sy1.8 & (SDSS J12168+0541) & (unk) & ... & Association? \\ 
\enddata
\tablecomments{ 
$^\dagger$ Extended.
}
\end{deluxetable*}
\end{longrotatetable}

\begin{deluxetable*}{c | cc | cc | cc | cc | cccc }[ht]
\tablecaption{Classes of cross-matched sources (\bat\ definition)
\label{tab:summary_bat}
}
\tablehead{
 \colhead{Source} & \multicolumn2c{Original} & \multicolumn2c{Matched point sources} & \multicolumn2c{Extended} & \multicolumn2c{ID-matched} & \multicolumn4c{Total-matched} 
\\ 
 \colhead{} & \colhead{\#} & \colhead{\%} & \colhead{\#} & \colhead{Firm } & \colhead{\#} & \colhead{Firm } & \colhead{\#} & \colhead{Firm} & \colhead{\#} & \colhead{\%} & \colhead{Firm \#} & \colhead{Firm \%}
} 
\startdata
                     Total &            1632 &              &            132 &                   115 &              31 &                             15 &             24 &                            21 &          187 &           &                         151 &                          \\
                     \hline
                Beamed AGN &             158 &             9.7 &             89 &                    89 &               1 &                              1 &             17 &                            17 &          107 &         57.2 &                         107 &                        70.9 \\
          Starburst galaxy &               1 &             0.1 &              1 &                     1 &               0 &                              0 &              0 &                             0 &            1 &          0.5 &                           1 &                         0.7 \\
            Seyfert galaxy &             827 &            50.7 &             10 &                     6 &               1 &                              0 &              4 &                             2 &           15 &          8.0 &                           8 &                         5.3 \\
                     LINER &               6 &             0.4 &              0 &                     0 &               0 &                              0 &              0 &                             0 &            0 &          0.0 &                           0 &                         0.0 \\
               Unknown AGN &             114 &             7.0 &              1 &                     0 &               2 &                              0 &              0 &                             0 &            3 &          1.6 &                           0 &                         0.0 \\
 Compact group of galaxies &               1 &             0.1 &              0 &                     0 &               0 &                              0 &              0 &                             0 &            0 &          0.0 &                           0 &                         0.0 \\
            Galaxy Cluster &              26 &             1.6 &              0 &                     0 &               0 &                              0 &              0 &                             0 &            0 &          0.0 &                           0 &                         0.0 \\
           Galactic Center &               1 &             0.1 &              1 &                     1 &               0 &                              0 &              0 &                             0 &            1 &          0.5 &                           1 &                         0.7 \\
                      HMXB &             108 &             6.6 &              5 &                     5 &               6 &                              0 &              0 &                             0 &           11 &          5.9 &                           5 &                         3.3 \\
                      LMXB &             109 &             6.7 &             10 &                     1 &               0 &                              0 &              1 &                             0 &           11 &          5.9 &                           1 &                         0.7 \\
                       XRB &               8 &             0.5 &              1 &                     1 &               1 &                              0 &              0 &                             0 &            2 &          1.1 &                           1 &                         0.7 \\
                    Pulsar &              25 &             1.5 &              5 &                     5 &              14 &                             12 &              2 &                             2 &           21 &         11.2 &                          19 &                        12.6 \\
                       SNR &               7 &             0.4 &              4 &                     4 &               2 &                              2 &              0 &                             0 &            6 &          3.2 &                           6 &                         4.0 \\
                      Nova &               6 &             0.4 &              0 &                     0 &               0 &                              0 &              0 &                             0 &            0 &          0.0 &                           0 &                         0.0 \\
                        CV &              75 &             4.6 &              1 &                     1 &               1 &                              0 &              0 &                             0 &            2 &          1.1 &                           1 &                         0.7 \\
            Symbiotic star &               4 &             0.3 &              0 &                     0 &               0 &                              0 &              0 &                             0 &            0 &          0.0 &                           0 &                         0.0 \\
                      star &              12 &             0.7 &              0 &                     0 &               1 &                              0 &              0 &                             0 &            1 &          0.5 &                           0 &                         0.0 \\
         Open star cluster &               1 &             0.1 &              0 &                     0 &               0 &                              0 &              0 &                             0 &            0 &          0.0 &                           0 &                         0.0 \\
           molecular cloud &               2 &             0.1 &              1 &                     0 &               0 &                              0 &              0 &                             0 &            1 &          0.5 &                           0 &                         0.0 \\
                        GC &               1 &             0.1 &              1 &                     1 &               0 &                              0 &              0 &                             0 &            1 &          0.5 &                           1 &                         0.7 \\
          Gamma-ray source &               1 &             0.1 &              0 &                     0 &               1 &                              0 &              0 &                             0 &            1 &          0.5 &                           0 &                         0.0 \\
           confused source &              10 &             0.6 &              1 &                     0 &               0 &                              0 &              0 &                             0 &            1 &          0.5 &                           0 &                         0.0 \\
                        U1 &              36 &             2.2 &              0 &                     0 &               0 &                              0 &              0 &                             0 &            0 &          0.0 &                           0 &                         0.0 \\
                        U2 &              55 &             3.4 &              0 &                     0 &               1 &                              0 &              0 &                             0 &            1 &          0.5 &                           0 &                         0.0 \\
                        U3 &              38 &             2.3 &              1 &                     0 &               0 &                              0 &              0 &                             0 &            1 &          0.5 &                           0 &                         0.0 \\
\enddata
\tablecomments{ 
Firm matches indicate sources with Flag being M or D, and do not include false-matched, unidentified, and ambiguous sources for safety.
The nearest source was used for the counterpart of the extended Fermi sources.
Here Seyfert galaxy includes all Seyfert 1 and 2 types.
}
\end{deluxetable*}

\begin{deluxetable*}{ c | cc | cc | cc | cc | cccc }[ht]
\tablecaption{Classes of cross-matched sources (4FGL-DR2 definition)
\label{tab:summary_fermi}
}
\tablehead{
 \colhead{Source} & \multicolumn2c{Original} & \multicolumn2c{Matched point sources} & \multicolumn2c{Extended} & \multicolumn2c{ID-matched} & \multicolumn4c{Total-matched} 
\\ 
 \colhead{} & \colhead{\#} & \colhead{\%} & \colhead{\#} & \colhead{Firm } & \colhead{\#} & \colhead{Firm } & \colhead{\#} & \colhead{Firm} & \colhead{\#} & \colhead{\%} & \colhead{Firm \#} & \colhead{Firm \%}
} 
\startdata
    Total  &            5788 &                &            132 &                   115 &              31 &                             15 &             24 &                            21 &          187 &           &                         151 &                          \\
    \hline
       BLL &            1190 &             21 &             32 &                    30 &               0 &                              0 &              3 &                             3 &           35 &         18.7 &                          33 &                        21.9 \\
      FSRQ &             730 &             13 &             43 &                    40 &               0 &                              0 &             10 &                            10 &           53 &         28.3 &                          50 &                        33.1 \\
       BCU &            1517 &             26 &             14 &                    13 &               0 &                              0 &              2 &                             2 &           16 &          8.6 &                          15 &                         9.9 \\
       AGN &              11 &           0.19 &              1 &                     1 &               0 &                              0 &              1 &                             1 &            2 &          1.1 &                           2 &                         1.3 \\
       RDG &              44 &           0.76 &              7 &                     7 &               1 &                              1 &              2 &                             2 &           10 &          5.3 &                          10 &                         6.6 \\
       SBG &               8 &           0.14 &              3 &                     3 &               0 &                              0 &              0 &                             0 &            3 &          1.6 &                           3 &                         2.0 \\
       SEY &               1 &          0.017 &              1 &                     1 &               0 &                              0 &              0 &                             0 &            1 &          0.5 &                           1 &                         0.7 \\
     NLSY1 &               9 &           0.16 &              1 &                     1 &               0 &                              0 &              0 &                             0 &            1 &          0.5 &                           1 &                         0.7 \\
      css  &               5 &          0.086 &              1 &                     1 &               0 &                              0 &              1 &                             1 &            2 &          1.1 &                           2 &                         1.3 \\
     ssrq  &               2 &          0.035 &              0 &                     0 &               0 &                              0 &              0 &                             0 &            0 &          0.0 &                           0 &                         0.0 \\
       GAL &               5 &          0.086 &              0 &                     0 &               2 &                              0 &              0 &                             0 &            2 &          1.1 &                           0 &                         0.0 \\
       SNR &              43 &           0.74 &              2 &                     2 &               7 &                              2 &              0 &                             0 &            9 &          4.8 &                           4 &                         2.6 \\
       PSR &             259 &            4.5 &             10 &                     7 &               0 &                              0 &              2 &                             2 &           12 &          6.4 &                           9 &                         6.0 \\
       PWN &              18 &           0.31 &              2 &                     1 &               9 &                              9 &              0 &                             0 &           11 &          5.9 &                          10 &                         6.6 \\
      spp  &              95 &            1.6 &              0 &                     0 &               5 &                              3 &              0 &                             0 &            5 &          2.7 &                           3 &                         2.0 \\
       BIN &               9 &           0.16 &              1 &                     1 &               0 &                              0 &              0 &                             0 &            1 &          0.5 &                           1 &                         0.7 \\
       HMB &               8 &           0.14 &              5 &                     5 &               0 &                              0 &              0 &                             0 &            5 &          2.7 &                           5 &                         3.3 \\
       LMB &               4 &          0.069 &              1 &                     1 &               0 &                              0 &              0 &                             0 &            1 &          0.5 &                           1 &                         0.7 \\
      glc  &              30 &           0.52 &              5 &                     1 &               0 &                              0 &              0 &                             0 &            5 &          2.7 &                           1 &                         0.7 \\
       SFR &               5 &          0.086 &              0 &                     0 &               1 &                              0 &              0 &                             0 &            1 &          0.5 &                           0 &                         0.0 \\
      NOV  &               1 &          0.017 &              0 &                     0 &               0 &                              0 &              0 &                             0 &            0 &          0.0 &                           0 &                         0.0 \\
 unidentified &            1794 &             31 &              3 &                     0 &               6 &                              0 &              3 &                             0 &           12 &          6.4 &                           0 &                         0.0 \\
\enddata
\tablecomments{ 
Firm matches indicate sources with Flag being M or D, and do not include false-matched, unidentified, and ambiguous sources for safety.
}
\end{deluxetable*}


\end{document}